\documentclass[10pt,prd,preprintnumbers,floatfix,aps,nofootinbib,showpacs,twocolumn,superscriptaddress,noshowpacs]{revtex4}

\usepackage{xspace}
\usepackage{lipsum}
\usepackage{relsize}
\usepackage{appendix}
\usepackage{amsmath,amstext,amssymb,amsfonts}
\usepackage{epsfig}
\usepackage{url}
\usepackage{nameref}
\usepackage{varioref}
\usepackage{hyperref}
\usepackage{cleveref}
\usepackage[normalem]{ulem} 

\usepackage{latexsym}
\usepackage{epsfig}
\usepackage{wasysym}
\usepackage{graphicx}
\usepackage{verbatim}
\usepackage{tikz} 
\usepackage{pgfplots}
\usepackage[export]{adjustbox}

\usepackage{bbold}

\usepackage[normalem]{ulem}

\newcommand{\CMU}{McWilliams Center for Cosmology, Department of Physics, Carnegie Mellon University, Pittsburgh, PA 15213, USA}

\newcommand{\plpeak}{\textsc{PLB}\xspace}
\newcommand{\plpl}{\textsc{PLPL}\xspace}
\newcommand{\plplpeak}{\textsc{PLPLB}\xspace}
\newcommand{\chieff}{\chi_{\rm{eff}}\xspace}

\begin{document}

\title{A new bump in the night: Evidence of a new feature in the binary black hole mass distribution at $70~M_{\odot}$ from gravitational-wave observations}

\email{imhernan@andrew.cmu.edu}
\author{Ignacio Maga\~na~Hernandez}
\affiliation{\CMU}
\author{Antonella Palmese}
\affiliation{\CMU}
\email{apalmese@andrew.cmu.edu}

\begin{abstract}
We analyze the confident binary black hole (BBH) detections from the third Gravitational-Wave Transient Catalog (GWTC-3) with an alternative mass population model in order to capture features in the mass distribution beyond the \textsc{Powerlaw + Peak} model. We find that the peak of a second power law characterizes the $\sim 30-35~ M_\odot$ bump, such that the data marginally prefers a mixture of two power laws for the mass distribution of binary components over a \textsc{Powerlaw + Peak} model with a Bayes Factor $\log_{10}\mathcal{B}$ of 0.24. This result may imply that the $\sim 30-35~ M_\odot$ feature represents the onset of a second population of BBH mergers (e.g. from a dynamical formation channel) rather than a specific mass feature over a broader distribution. When an additional Gaussian bump is allowed within our power law mixture model, we find a new feature in the BH mass spectrum at $\sim65-70~M_\odot$ ($\log_{10}\mathcal{B}$ = 0.29 compared to \textsc{Powerlaw + Peak}). This new feature may be consistent with hierarchical mergers, and constitute $\sim3\%$ of the BBH population. This model also recovers a maximum mass of $58^{+32}_{-14}~M_\odot$ for the second power law, consistent with the onset of a pair-instability supernova mass gap.
\end{abstract}

\date{\today}
\maketitle

\section{Introduction} \label{sec:Intro}

Since the first detection of a binary black hole (BBH) merger in gravitational waves (GW)~\cite{LIGOScientific:2016aoc}, the field of GW astronomy has entered the era of population studies. During the course of the first three observing runs conducted by LIGO and Virgo, 69 high significance BBH mergers have been detected \citep{gwtc3_data}, and hundreds are expected to be identified throughout the ongoing fourth LIGO-Virgo-KAGRA (LVK) observing run.

Multiple formation channels have been proposed to explain the observed population of compact objects beyond the standard isolated binaries scenario, including formation pathways requiring dynamical interactions in dense star clusters~\citep{Rodriguez_2015,rodriguez19}, in the disks of Active Galactic Nuclei (AGN)~\citep{2012MNRAS.425..460M,Antonini_2016,2018ApJ...866...66M} or in the center of dwarf galaxies \citep{Conselice_20,palmese_conselice}, black holes from metal-free Population III stars (Pop III)~\citep{Kinugawa2014MNRAS.442.2963K,Hartwig_2016}, and primordial black holes~\citep{Carr:1974nx,Clesse:2016ajp,Tsai_2021}. For in depth reviews, see~\citep{Mapelli_2021,Mandel_2022}.

It is becoming increasingly clear that there is no single dominating channel responsible for the entire population, but rather a combination of formation mechanisms which can explain the various features and correlations observed in the data \citep{Zevin_2021,2023ApJ...955..127C,ray2024searchingbinaryblackhole}.

Multiple options for fitting the BBH mass distribution (and more broadly other population properties) have been proposed. The second and third Gravitational-Wave Transient Catalog (GWTC) fiducial analyses include a broken power law or a power law plus peak \citep{LIGOScientific:2020kqk,gwtc3_sensitivity}. Recent studies have also modeled the distribution with physics-motivated models \citep{Baxter_2021} or non-parametric approaches \citep{Mandel_2016,gwtc3_sensitivity,Edelman:2022ydv,ray2024searchingbinaryblackhole,MaganaHernandez:2024uty,Farah:2024xub}.

Evidence for multiple BH populations includes the presence of a peak in the primary mass distribution at $\sim30-35~M_\odot$, which is seemingly too narrow or too unrealistically located to to be explained by various formation channels~\citep{Briel:2022cfl,Golomg_2023,Hendriks:2023yrw}. Furthermore, BBH from Pop III stars, which may be the seeds of of galaxies' central black holes still residing in ultra-dwarf galaxies in the nearby Universe, are also predicted to peak around these values~\citep{Kinugawa2014MNRAS.442.2963K,Conselice_20,palmese_conselice}. Motivated by these theories, we explore whether the $\sim30-35~M_\odot$ excess can be explained by a second power law rather than a peak over a broader power law by using the confidently detected BBH observations from the GWTC-3 catalog ~\citep{gwtc3_data}).

Given that we are considering a population model for a formation channel which is possibly of dynamical origin (see~\cite{Gerosa:2021mno} for a review), we also explore the presence of an additional high mass hierarchical merger population \citep{LIGOScientific:2020ufj,Kimball_2020,Kimball:2020qyd} that could originate from the merger of two black holes each formed from the merger of another two compact objects in the $30-35~M_\odot$ peak of the power law.

In \S\ref{sec:method} we present our method, in \S\ref{sec:Results} we show our results on GWTC-3, in \S\ref{sec:validation} we present validation tests using simulated GW events, and discuss spin measurements in \S\ref{sec:spin}. In \S\ref{sec:astro} we discuss astrophysical implications of our findings, and in \S\ref{sec:conclusion} we outline our summary and conclusions. Uncertainties throughout this work are at the 90\% Credible Interval (CI).

\section{Methods}\label{sec:method}
We use hierarchical Bayesian population inference to simultaneously infer the parameters describing the mass and redshift distributions of BBH mergers with parametric models. We provide a brief summary below and we refer the reader to review
articles for an in depth description of the framework \citep{Mandel_2019, Thrane_2019, Vitale_2021}. 

\subsection{Hierarchical Bayesian Inference}
The BBH observations provide us with an estimate of their primary mass $m_1$ and secondary mass $m_2$ components (in the source frame), as well as the measured redshift of the source $z$.

We can write the number density of BBH events as
\begin{equation}
     \frac{dN(m_1, m_2, z | \Lambda)}{dm_1 dm_2 dz} \propto \frac{dV_c}{dz}\bigg(\frac{T_\mathrm{obs}}{1+z}\bigg) \mathcal{R}_0(1+z)^\kappa p(m_1, m_2 | \Lambda)
\end{equation}
\noindent
where $\Lambda$ denotes the hyperparameters describing the population distribution $p(m_1, m_2 | \Lambda)$ of BBH masses. Here, $dV_c/dz$ is the differential uniform-in-comoving volume element, $T_\mathrm{obs}$ is the total observation time, the factor of $1/(1+z)$ converts source-frame time to detector-frame time, and $\mathcal{R}_0$ is the local BBH merger rate at $z=0$. The BBH population is modeled through the normalized mass distribution $p(m_1, m_2, | \Lambda)$ as well as its redshift evolution, which we parameterize by a power law proportional to $(1+z)^\kappa$ as in previous studies \citep{Fishbach_2018,KAGRA:2021duu}. 

Given a set of $N_\mathrm{obs}$ gravitational wave observations $\{d_i\}$, we can calculate the posterior on $\Lambda$ following~\citep{Farr_2019} and \citep{Mandel_2019}:
\begin{align} \label{eqn:posterior}
\begin{split}
    p\left(\Lambda | \{d_i\} \right)  &\propto \frac{p(\Lambda)}{\beta(\Lambda)^{N_\mathrm{obs}}} \prod_{i=1}^{N_\mathrm{obs}} \Bigg[ \int \mathcal{L}\left(d_i | m_1, m_2, z \right) \\
    & \times \frac{dN(m_1, m_2, z | \Lambda)}{dm_1dm_2dz} dm_1 dm_2 dz \Bigg],
\end{split}
\end{align}
\noindent
where $\mathcal{L}(d_i|m_1, m_2, z)$ is the single-event likelihood function for each event, and $\beta(\Lambda)$ is the detectable fraction of sources corresponding to a population determined by the population hyperparameters $\Lambda$. 

Following~\cite{Tiwari_2018} and~\cite{Farr_2019}, we estimate $\beta(\Lambda)$ by assuming that it will follow a normal distribution (i.e. $\beta(\Lambda) \sim \mathcal{N}(\Lambda|\mu, \sigma)$), with $\mu$ as the importance sample estimate of $\beta$ and its associated uncertainty $\sigma^2 = \mu^2/N_\mathrm{eff}$. Here $N_\mathrm{eff}$ is the number of effective samples which contribute to the estimate of $\beta$ given a population model described by $\Lambda$. As in Ref.~\cite{Farr_2019}, we make sure that $N_\mathrm{eff} > 4 N_\mathrm{obs}$ to provide an unbiased estimate of $\beta$. Practically, we estimate the detectable fraction $\beta(\Lambda)$ by using the LVK's injection campaign of BBH events simulated from a broad BBH population, injected into real detector data, and then searched for using the same analysis pipelines which found the events in GWTC-3~\citep{KAGRA:2021duu}. 

Finally, we marginalize over the local BBH merger rate $\mathcal{R}_0$ using a log-Uniform prior \citep{Farr_2019,LIGOScientific:2020kqk}. We also approximate the integral over the individual event likelihoods in Eq.~\ref{eqn:posterior} with importance sampling over $N_i$ single-event posterior samples generated by the LVK from single event parameter estimation (PE) analyses with prior choices corresponding to $\pi(m_1, m_2, z) \propto d_L^2 (1+z)^2 \frac{dd_L}{dz}$ \cite{KAGRA:2021duu}. In Sec.~\ref{sec:spin}, we analyze additional spin information for each merger encoded in their $\chieff$ measurement (a mass-weighted, aligned, \emph{effective spin}, see Sec.~\ref{sec:spin} for its definition). Therefore, we compute the default PE prior $\pi(m_1, m_2, z, \chieff)$ according to \cite{Callister:2021gxf} using the \texttt{gw-distributions} library\footnote{publicly available at \url{https://git.ligo.org/reed.essick/gw-distributions}}.

\subsection{Model Comparison}
Given the estimated hyper-parameter posterior distribution on $\Lambda$ for a given model $M$, we can compute the evidence for this model as: 
\begin{equation}
Z_M = \int p(\{d_i\}, \Lambda | M) d\Lambda = \int p(\{d_i\} | \Lambda, M) p(\Lambda, M)
\end{equation}

If we have a competing model $N$, then the ratio of their evidences, the Bayes factor, provides us with a figure of merit to perform model comparison \citep{Jefferys1961}. Here we use, 
\begin{equation}
    \log_{10}\mathcal{B}^{M}_{N} = \log_{10}\left(\frac{Z_M}{Z_N} \right)
\end{equation}
where we have decided to work in log base 10 units for convenience. In general if $\log_{10}\mathcal{B}^{M}_{N} > 0$, then model $M$ is favored over model $N$. Likewise, if $\log_{10}\mathcal{B}^{M}_{N} < 0$, the opposite is true. If $\log_{10}\mathcal{B}^{M}_{N} \approx 0$, then it is inconclusive if one model is preferred over the other, and therefore both are equally good models at representing the underlying population given available data. 

\subsection{Population Models}
\label{sec:models}
Following~\cite{farah:2023swu}, we model the population mass distribution as a symmetric distribution $p(m|\Lambda)$ on both the primary and secondary mass with a pairing function $f(m_1,m_2|\Lambda)$. Here $\Lambda$ are the hyperparameters describing the mass models. That is, we consider the following family of models,
\begin{equation}
    p(m_1,m_2|\Lambda) \propto p(m_1|\Lambda)p(m_2|\Lambda)f(m_1,m_2|\Lambda).
\end{equation}
More specifically, we consider a pairing function which depends on the mass ratio $q=m_2/m_1$, or
\begin{equation}
    f(m_1,m_2|\beta) = \left( \frac{m_2}{m_1}\right)^\beta.
\end{equation}

In this work, we consider the following family of models for $p(m|\Lambda)$ as a mixture of power laws $p_{{\rm{pl}}}$ and Gaussian $p_{{\rm{g}}}$ component probability distributions,
\begin{equation}
    p(m|\Lambda) = \sum_i f_{{\rm{pl}},i} p_{{\rm{pl}},i}(m|\Lambda) + \sum_j f_{{\rm{g}},j} p_{{\rm{g}},j}(m|\Lambda)
\end{equation}
where the mixture weights must add up to one or, $\sum_i f_{{\rm{pl}},i}  + \sum_j f_{{\rm{g}},j} = 1$.

More explicitly, we consider a mixture of two power laws and a single Gaussian component, and therefore we write,
\begin{widetext}
\begin{equation}\label{eqn:model}
    p(m|\Lambda) = \sum_{i=1}^{2} f_{{\rm{pl}},i}S(m|m_{{\rm{min}},i}, \delta m_{{\rm{min}},i})S(m|m_{{\rm{max}},i}, \delta m_{{\rm{max}},i})p_{{\rm{pl}},i}(m|m_{{\rm{min}},i},m_{{\rm{max}},i},\alpha_i) + f_{{\rm{g}}}p_{{\rm{g}}}(m|\mu_m, \sigma_m)
\end{equation} 
\end{widetext}
where $\left( m_{{\rm{min}},i},m_{{\rm{max}},i},\alpha_i \right)$ correspond to the minimum mass, maximum mass, and power law slope for the $i$th power law component, respectively. $\mu_m$ and $\sigma_m$ represent the mean and standard deviation for the Gaussian component. The $S$ functions smooth each of the power law distribution components at both their lower and higher ends with smoothing scales $\delta m_{{\rm{X}},i}$ for $X = \{\rm{min},\rm{max}\}$. As before, the mixture weights add up to unity, i.e. $f_{{\rm{pl}},1} + f_{{\rm{pl}},2} + f_{{\rm{g}}} = 1$.

With this in mind, we consider the following population models for the mass distribution of BBH mergers. First, if we set $f_{{\rm{pl}},2} = 0$ we obtain the model with a single power law and a single Gaussian component, denoted by \textsc{Powerlaw+Bump} (\plpeak). This corresponds to the model described in detail in \cite{farah:2023swu}, however we do not refer to it as \textsc{Powerlaw+Peak} to avoid confusion with the model in \cite{gwtc3_data}. Second, we set the Gaussian component mixture weight to zero, i.e. $f_{{\rm{g}}} = 0$, and consider a model described by a mixture of two power laws, dubbed here as \textsc{Powerlaw+Powerlaw} (\plpl). Finally, we also consider the three-distribution mixture model described in Eq.~\ref{eqn:model} and denote it by \textsc{Powerlaw+Powerlaw+Bump} (\plplpeak).

\section{Results} \label{sec:Results}
In this section, we show the inferred population distributions for the masses of BBH mergers using the models we described in Section~\ref{sec:models}. Following the same GWTC-3 \citep{KAGRA:2023pio} population analysis choices \citep{KAGRA:2021duu}, we only use the BBH events which pass a 1 per year False Alarm Rate (FAR) threshold. Therefore, we analyze a total of 69 confident BBH detections and exclude population outliers such as GW190814~\citep{LIGOScientific:2020zkf} from our analysis. For each event used in our analysis, we make use of the parameter estimation (PE) samples released by LVK \citep{gwtc3_data}. We also make use of the publicly available LVK GWTC-3 sensitivity injection campaign \citep{gwtc3_sensitivity} to estimate the detection fraction $\beta(\Lambda)$ due to selection effects.

\begin{figure*}
\begin{center}
\includegraphics[width=\textwidth]{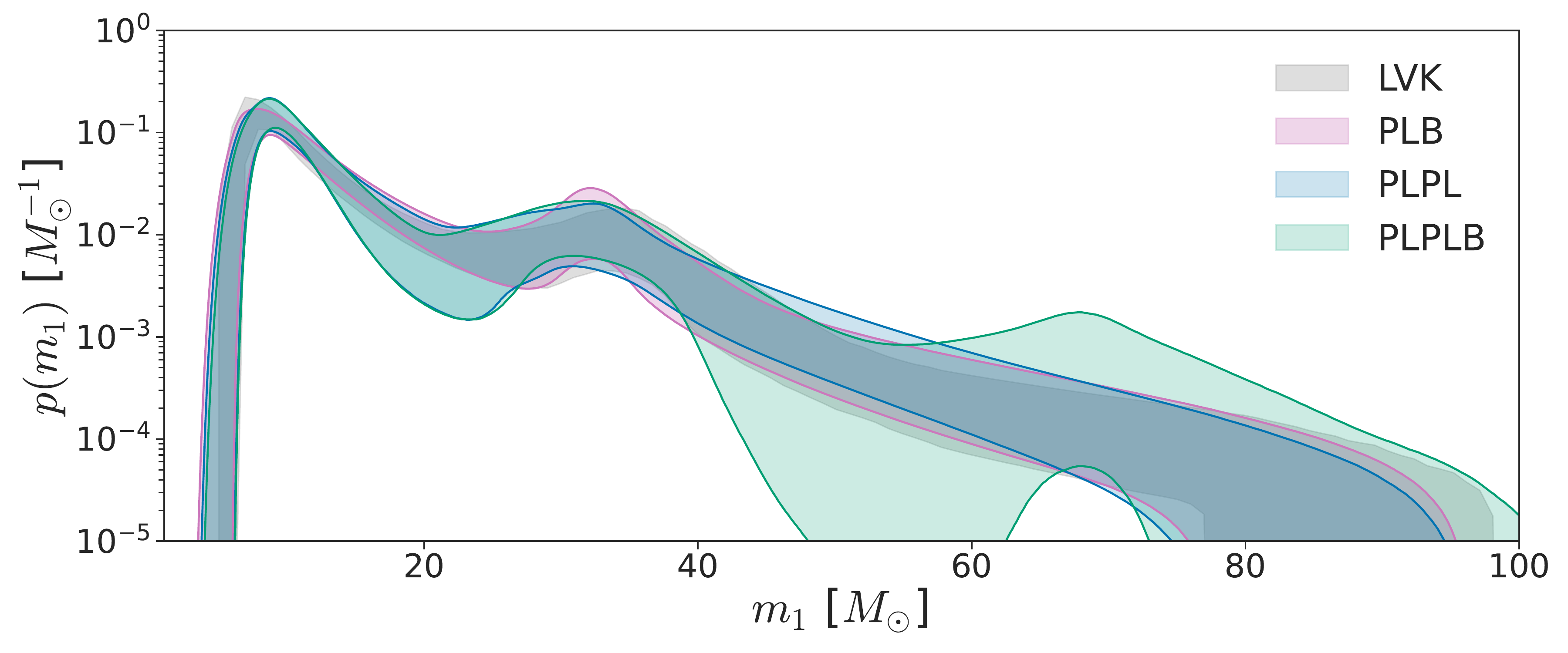}
\includegraphics[width=\textwidth]{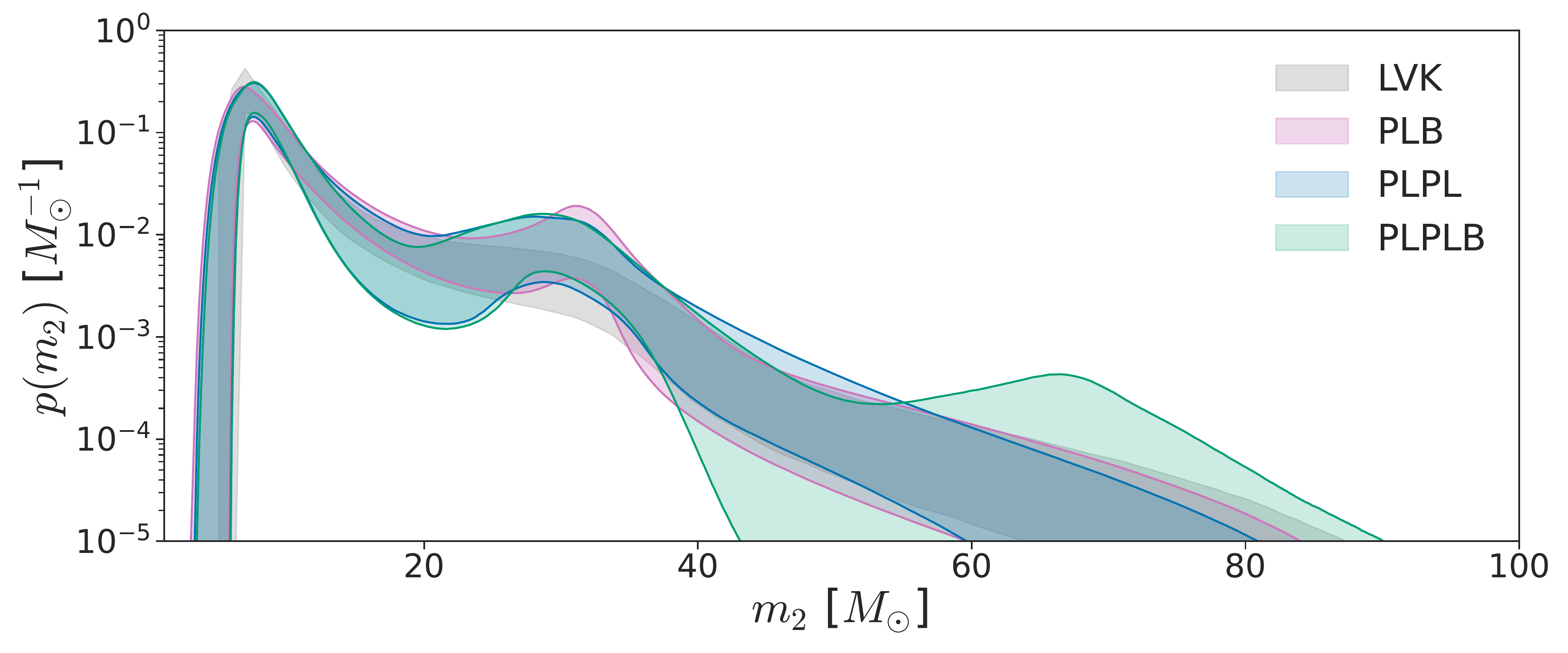}
\caption{\label{fig:gwtc3} Marginalized primary (top panel) and secondary (lower panel) mass distributions inferred with the \plpeak (pink), \plpl (blue) and \plpeak (green) models. For reference, we also show the GWTC-3 \textsc{Powerlaw+Peak} constraints in gray from \cite{KAGRA:2021duu}. }
\end{center}
\end{figure*}

We show the main results of this paper in Fig.~\ref{fig:gwtc3}. We plot the marginal posterior distributions for both the primary and secondary masses for all models considered in this work. We also show the LVK GWTC-3 \textsc{Powerlaw+Peak} result for reference. We note that this distribution differs from \plpeak because it parameterizes $(m_1,q)$ rather than $(m_1,m_2)$, and consequently sets different distribution shapes on both $m_1$ and $m_2$, with the peak present only in the primary mass distribution \citep{farah:2023swu}.

First, we note that the \plpl model from a qualitative perspective appears to be a good fit to the GWTC-3 data since the 90\% confidence interval on the marginal mass distributions are consistent with \textsc{Powerlaw+Peak}. More quantitatively, we can compute the Bayes factor comparing the \plpl model to the \plpeak model to obtain an estimate of how well the two models represent the GWTC-3 population with respect to each other. We obtain $\log_{10}\mathcal{B^{\plpl}_{\plpeak}} = +0.24$, and determine that the \plpl model describes the GWTC-3 observations similarly well to the \plpeak model, if not marginally better. In other words, the Gaussian bump at around $30-35~M_{\odot}$ could be described by a power law which is steeper than that recovered with the \plpeak model, turning on at a minimum mass of $m_{{\rm{min}},2} \sim 18~M_{\odot}$ and maximum mass $m_{{\rm{max}},2} \sim 45-100 ~M_{\odot}$ (at a different location depending on which model is chosen between \plpl and \plplpeak), potentially changing the astrophysical interpretation of this population. 

\begin{figure}
\begin{center}
\includegraphics[width=0.45\textwidth]{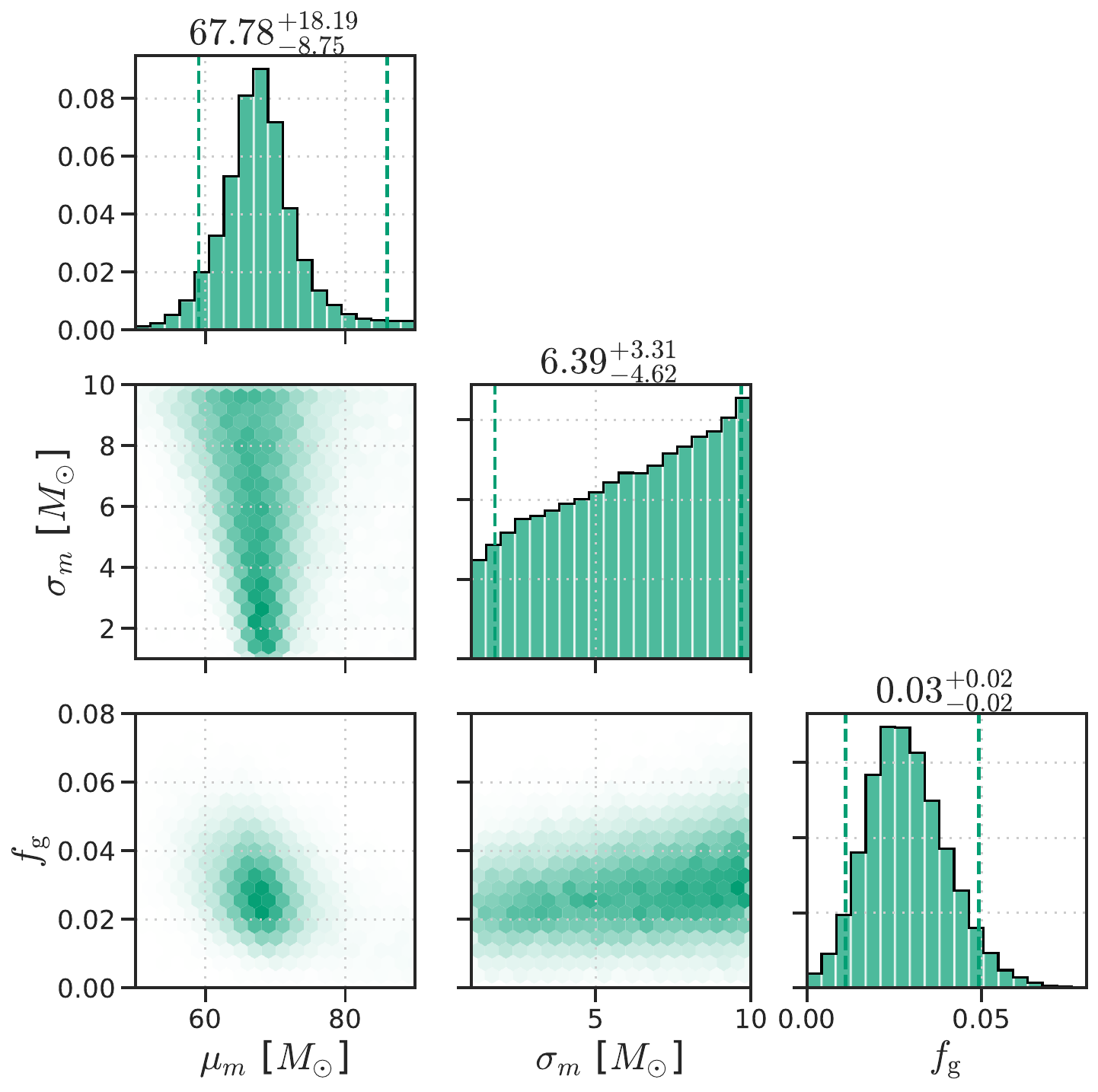}
\caption{\label{fig:bump} Marginalized posterior distribution on the location $\mu_m$, width $\sigma_m$ and mixture weight $f_{\rm{g}}$ for the \plplpeak model. We find the contribution to the overall population is 1-4\%.} 
\end{center}
\end{figure}

Second, we note that the more complex model, namely the \plplpeak model, appears to find a feature at approximately $70~M_{\odot}$. In Fig.~\ref{fig:bump}, we show the marginalized posterior distribution on $\{\mu_m, \sigma_m, f_{\rm{g}}\}$. Under the \plplpeak model, we find consistency on the measured power law slope and minimum/maximum masses for the first power law compared to the \plpl model. As for the second power law, we find a consistent measurement for $m_{{\rm{min}},2}$ but find that its maximum mass shifts to a lower mass $m_{{\rm{max}},2} \sim~50 M_{\odot}$ with a well defined mode compared to being preferentially larger than $\sim 70~M_{\odot}$. We find broad but consistent posteriors on $\alpha_2$ and the smoothing scales. The presence of the $\sim70~M_\odot$ feature is robust with respect to the prior on the position $\mu_m$ of the Gaussian peak. For more details, we refer the reader to Appendix~\ref{app:appendix} for corner plots showing the posterior distributions of each model considered in this work.

Again, we compute a Bayes factor for this model against \plpeak and find $\log_{10}\mathcal{B^{\plplpeak}_{\plpeak}} = +0.29$. Our results correspond to a Bayes factor of our \plplpeak model compared to \plpl of $\log_{10}\mathcal{B^{\plplpeak}_{\plpl}} = +0.05$, which is reasonable considering the additional complexity of the model. Therefore, we note that the differences in our estimates are not large enough to rule out the \plplpeak model and argue that more data is necessary to conclusively claim whether \plpl is preferred over \plplpeak. On the other hand, the \plpl model shows degeneracies with $m_{{\rm{max}},2}$ and $f_{\rm{pl},1}$ demonstrating the necessity for an extended model, therefore motivating the \plplpeak extended model (see Appendix~\ref{app:appendix} for more details and for full posterior distributions on every model considered in this paper). Most interestingly, it seems that the bimodality present in these two parameters posteriors stems from the two cases where the 65-70$~M_\odot$ population is described by PL1 or PL2, so that $m_{{\rm{max}},2}$ is either above or below 65-70$~M_\odot$, but shows very little support around this mass range. In other words, the presence of a population around this mass range does not allow for a mass gap to be present there. This bimodality is broken in the \plplpeak model.

\begin{figure}
\begin{center}
\includegraphics[width=0.45\textwidth]{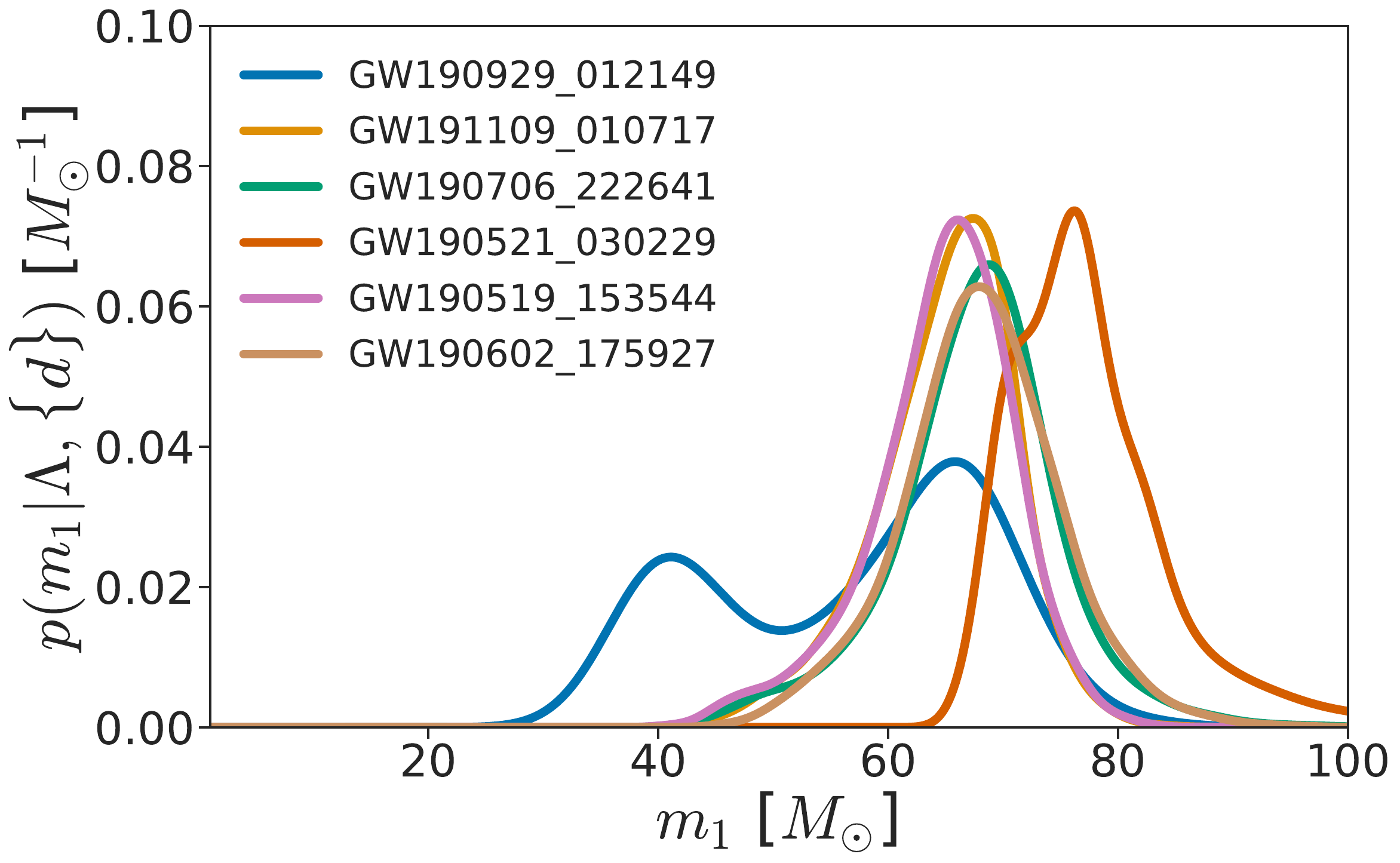}
\includegraphics[width=0.45\textwidth]{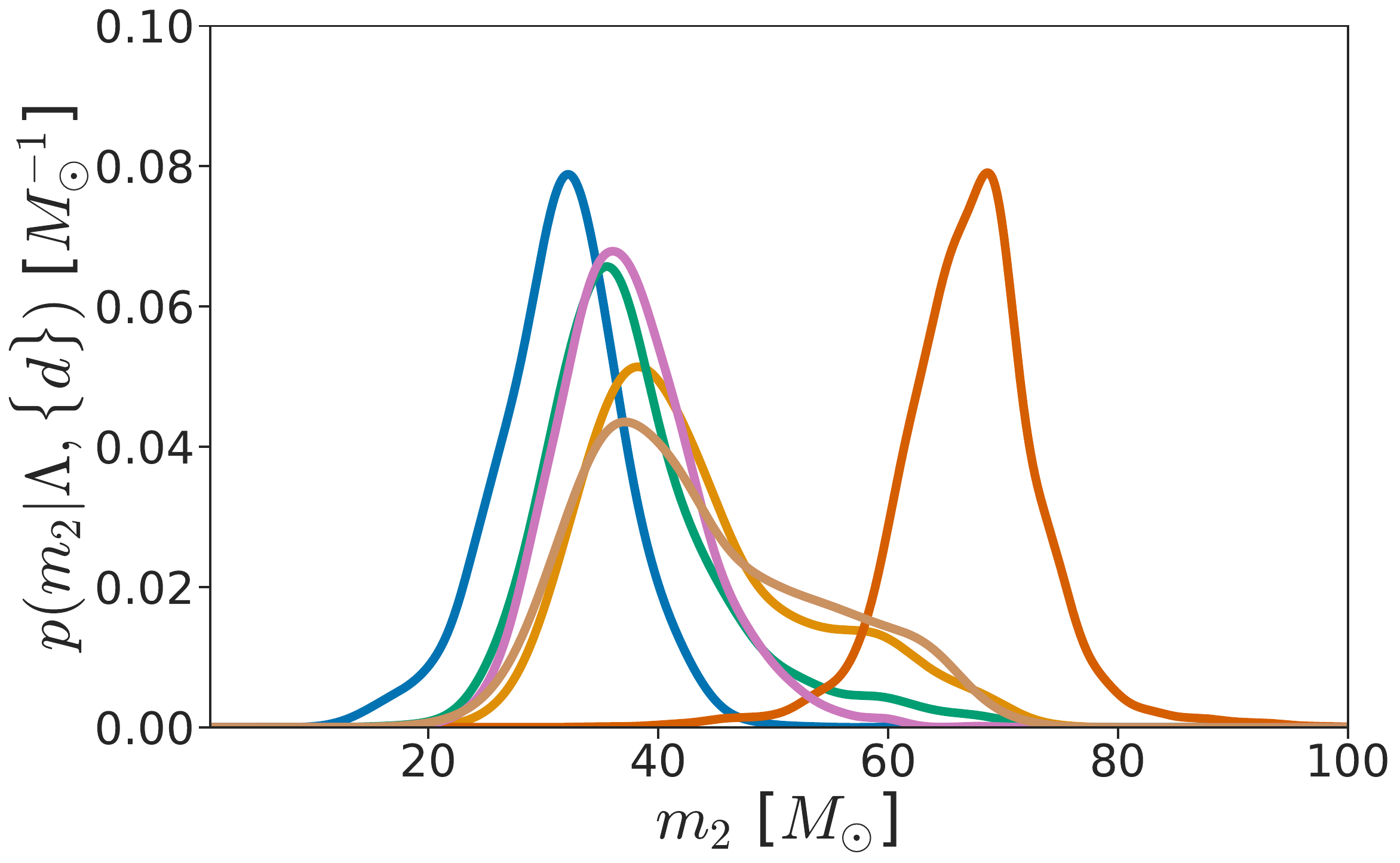}
\caption{\label{fig:events} Marginalized \plplpeak population-informed primary (top panel) and secondary (lower panel) mass posteriors for events in GWTC-3 that are consistent to have a component BH with mass larger than 55$~M_\odot$. }
\end{center}
\end{figure}

An interesting question to answer is: which events in GWTC-3 contribute mostly to the high mass bump? Using our \plplpeak population fit, we re-weight the single event source frame masses and redshift posterior samples to match its ensemble properties while marginalizing over its uncertainty~\citep{Moore_2021}. With the \plplpeak population-informed single event masses, we look for events with $m_1$ or $m_2$ larger than 55$~M_\odot$ and where the number of posterior samples above our mass threshold is at least 50\%. We find six events in GWTC-3 that satisfy this criterion and show the marginalized and population-informed posterior distributions on $m_1$ and $m_2$ in Fig.~\ref{fig:events}. As expected, GW190521 is one of those events, and both the primary and secondary contribute to the high mass bump. The primary masses of five other events contribute significantly to the peak, implying that the results found for the \plplpeak model are not only driven by a single outlier event. 

Finally, if we extend the \plplpeak model to have a second Gaussian component between $[20-50]~M_{\odot}$ (as in the \plpeak model), the feature at $\sim30-35~M_\odot$ is still dominated by the second power law and the conclusions of this study are not changed.

\section{Validation}\label{sec:validation}
\begin{figure}
\begin{center}
\includegraphics[width=0.48\textwidth]{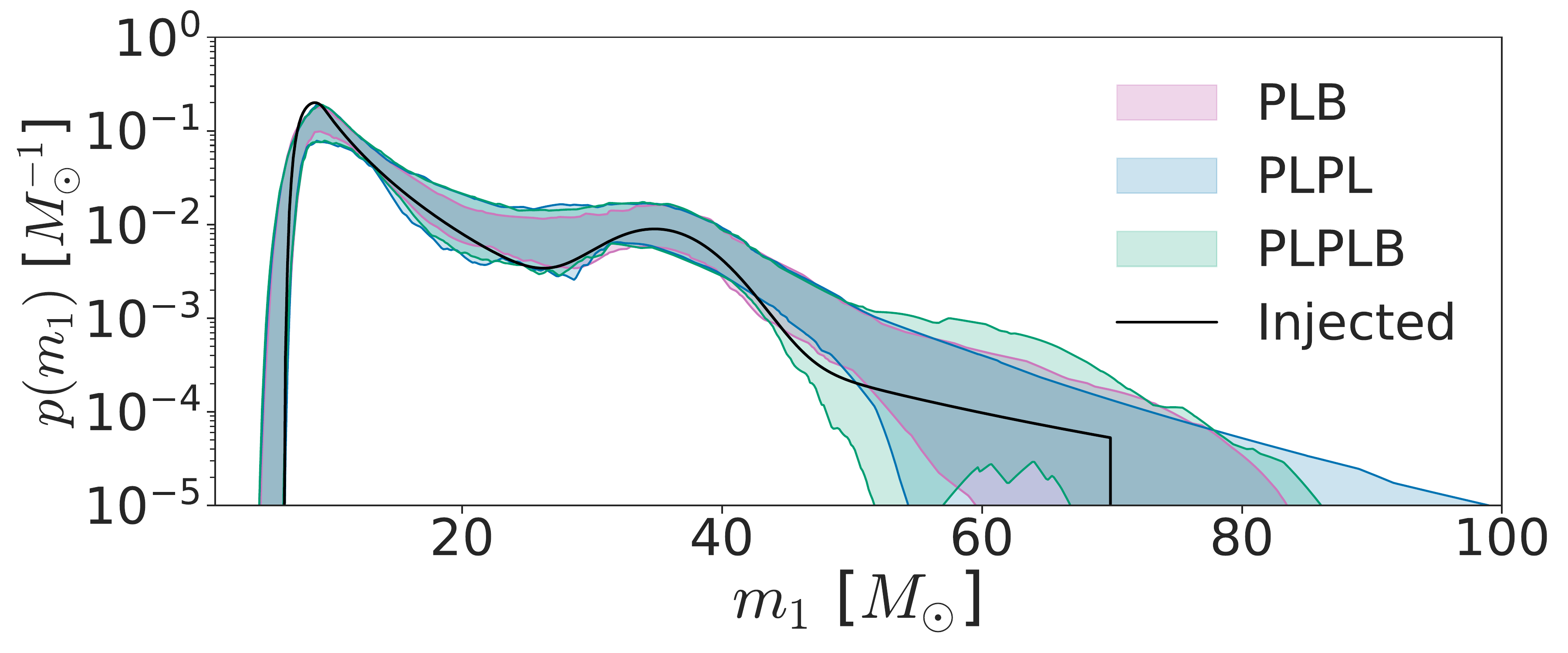}
\includegraphics[width=0.48\textwidth]{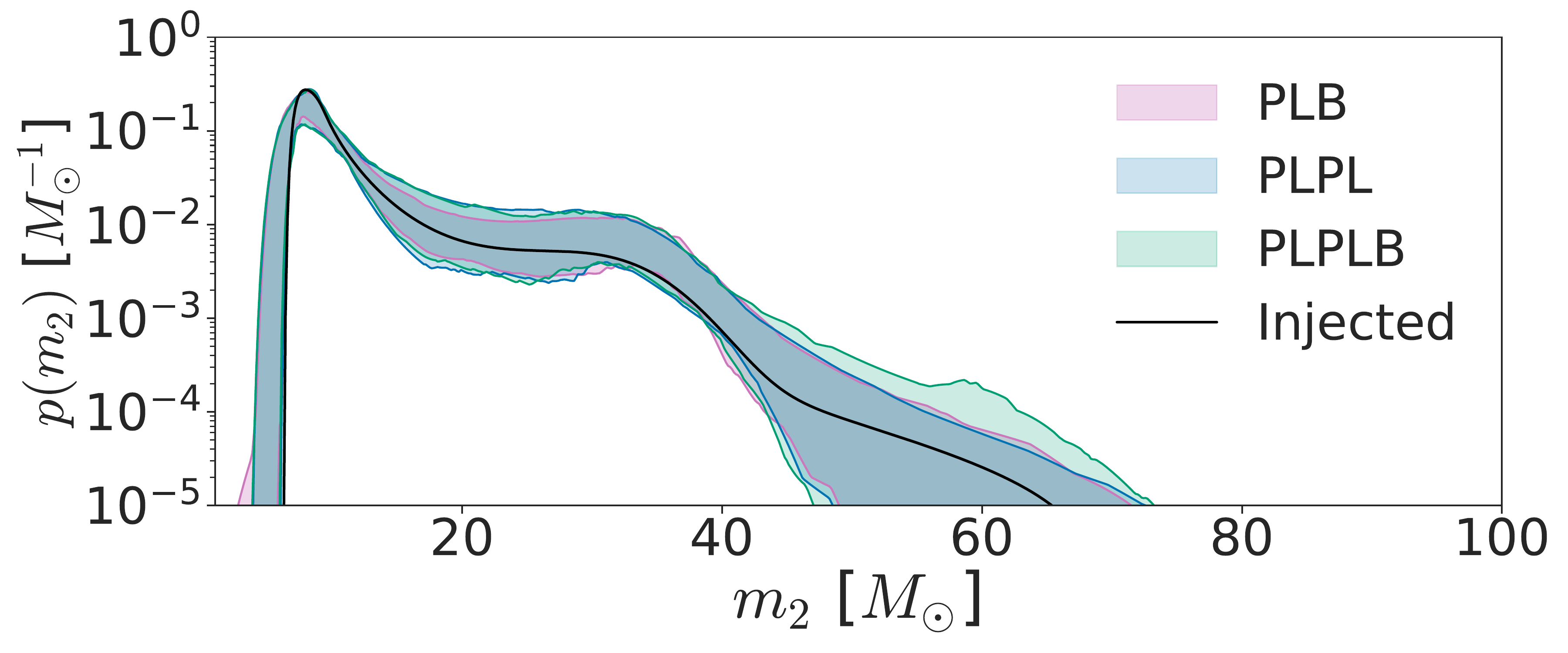}
\caption{\label{fig:sims} Primary (top panel) and secondary (lower panel) component mass distributions (90\% confidence interval) from the median mass distribution reconstructions for each of the 100 simulated GWTC-3-sized catalogs described in Section~\ref{sec:validation} for the \plpeak (pink), \plpl (blue) and \plpeak (green) models. For reference, we show the simulated GWTC-3 \textsc{Powerlaw+Peak} like population as the solid black line.}
\end{center}
\end{figure}

We validate our results using a large set of mock GW observations drawn from a known fiducial population model. Specifically, we want to test whether a \plpl or \plplpeak model can spuriously fit the data given a BBH population which actually follows a \plpeak model. We draw simulated BBH events using the default LVK \textsc{Powerlaw + Peak} model~\citep{Talbot:2018cva} for the primary mass distribution. Mass ratios are drawn from a power law with slope $\beta$, and we allow the overall merger rate to evolve with redshift as a power law in $(1+z)^\kappa$. Our simulations are constructed to closely follow the inferred GWTC-3 population parameters from~\cite{KAGRA:2021duu}, hence the fiducial choice for the hyper-parameter values for the simulated population described in~\cite{MaganaHernandez:2024uty}. For reference, the marginalized component mass distributions for our simulated population are shown in Figure~\ref{fig:sims}. 

Subsequently, we assign observed values to detectable BBH systems, i.e. those which meet our selection criteria and calibrate our measurement uncertainty to what is expected for BBH mergers with the Advanced LIGO design sensitivity~\citep{Fishbach_2018,Fishbach_2020,KAGRA:2013rdx}. We also simulate a corresponding set of detectable BBH systems from a broad fiducial population to take selection effects into account~\citep{Tiwari_2018,Farr_2019}.

\begin{figure}
\begin{center}
\includegraphics[width=0.48\textwidth]{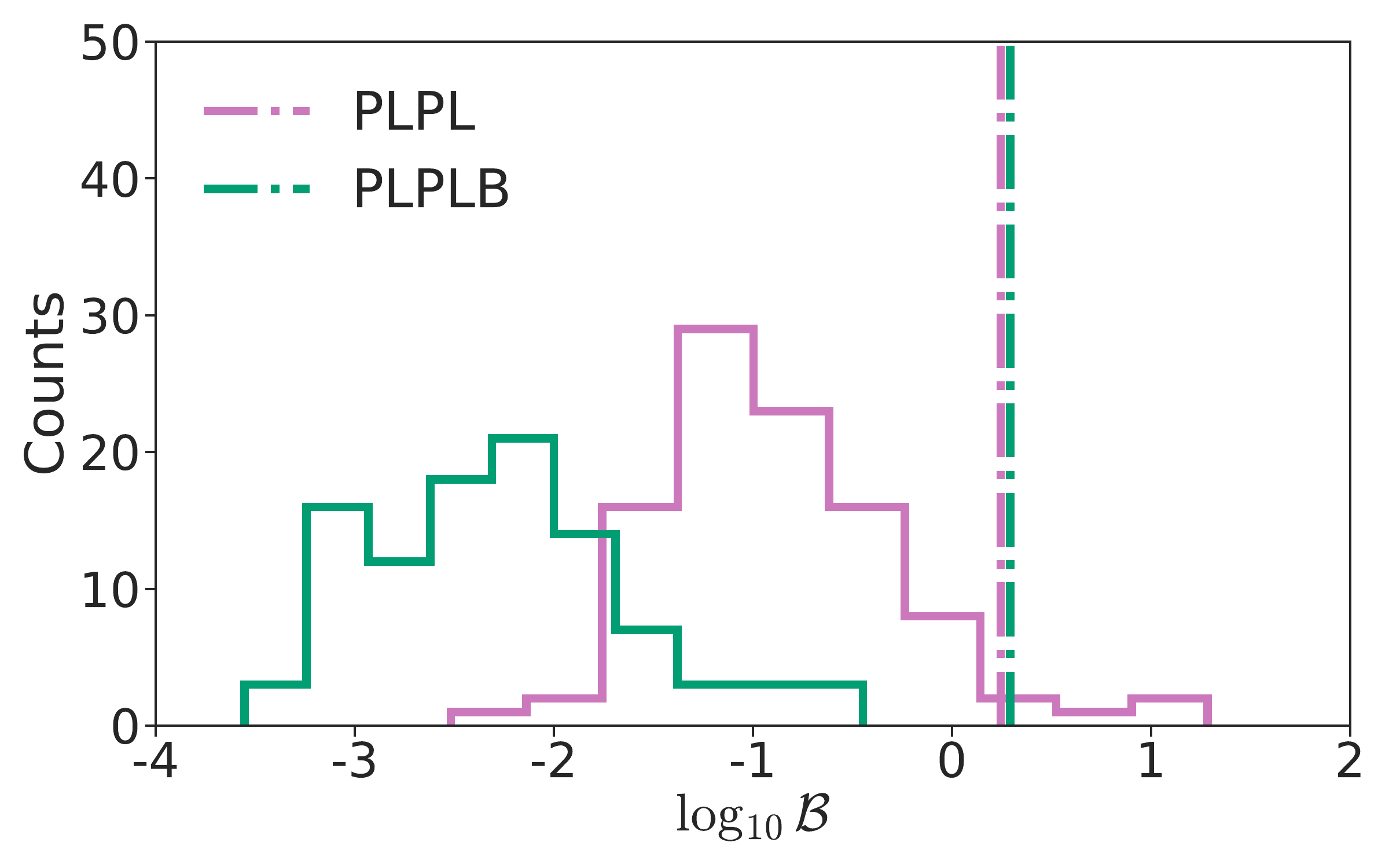}
\caption{\label{fig:histograms} In pink (green) we show the distribution of $\log_{10}\mathcal{B^{\plpl}_{\plpeak}}$ ($\log_{10}\mathcal{B^{\plplpeak}_{\plpeak}}$) for the simulated GWTC-3-sized catalogs described in Sec.~\ref{sec:validation}. The vertical lines show the corresponding GWTC-3 results.}
\end{center}
\end{figure}

The goal of our validation study is to understand the distribution of $\log_{10}\mathcal{B}$ obtained on GWTC-3-sized catalogs. To do this, we use 69 mock BBH observations (drawn from our simulated catalog) and repeat this procedure so that we have 100 independent realizations in total. We then perform hierarchical population inference on each of our GWTC-3-sized catalogs with each of the population models considered in this work. We then compute the corresponding $\log_{10}\mathcal{B^{\plpl}_{\plpeak}}$ and $\log_{10}\mathcal{B^{\plplpeak}_{\plpeak}}$ factors for each catalog realization. In Fig.~\ref{fig:histograms}, we show the distribution of $\log_{10}\mathcal{B}$ for both the \plpl and \plplpeak models (compared to \plpeak) and show the corresponding GWTC-3 results from Section~\ref{sec:Results} for reference. 

Concerning the \plpl model, there are some simulated realizations which show a preference over the \plpeak model, i.e., $\log_{10}\mathcal{B^{\plpl}_{\plpeak}}>0$ even when our simulated population was constructed with \textsc{Powerlaw+Peak} as the fiducial model. We find that $\log_{10}\mathcal{B^{\plpl}_{\plpeak}}>0$ in 7\% of our simulations. This shows that given the current size of the GWTC-3 catalog and its measurement uncertainty, we cannot confidently claim that \plpl is a preferred model over \plpeak. This is in agreement with the corresponding $\log_{10}\mathcal{B^{\plpl}_{\plpeak}}$ for the GWTC-3 data, as can be seen from the overlap with the estimated distribution of Bayes factors. 

More interestingly, we see that for the \plplpeak model, given our assumptions on the simulation inputs, it is unlikely to reproduce a $\log_{10}\mathcal{B^{\plplpeak}_{\plpeak}}$ factor consistent with the one measured with GWTC-3. We also note that for all the simulated realizations, we always find $\log_{10}\mathcal{B^{\plplpeak}_{\plpeak}}<0$. This means that \plplpeak is not a preferred representation of the underlying population if it is drawn from \textsc{Powerlaw+Peak}. Thus, our results and validation simulations potentially hint that a sub-population of BBH mergers at around $70~M_{\odot}$ is real, rather than a spurious feature due to the small number statistics and the specific representation we observed with GWTC-3.

As a secondary check, in Fig.~\ref{fig:sims}, we plot the the 90\% confidence interval constructed from the median component mass distribution reconstructions for each of the simulated GWTC-3-sized catalogs using the mass models considered in this work. We first note that \plpeak, \plpl, and \plplpeak are all able to recover the underlying \textsc{Powerlaw+Peak} population. 

For the \plplpeak model, we are most interested in determining how often and with which amplitude the Gaussian component fits a non-existent feature in the range of $60-80~M_\odot$. This is done to understand how sensitive our model is to the potentially small number statistics for the GWTC-3 catalog realizations and how, for some realizations, we might get clusters of events in the $60-80~M_\odot$ mass range when no sub-population is present. We see that this can happen, however, conditional on our simulation assumptions we see that we cannot reproduce our \plplpeak GWTC-3 results, e.g. the amplitude of the Gaussian feature that we measure is a lot stronger than what we see in our simulations.

\begin{figure}
\begin{center}
\includegraphics[width=0.48\textwidth]{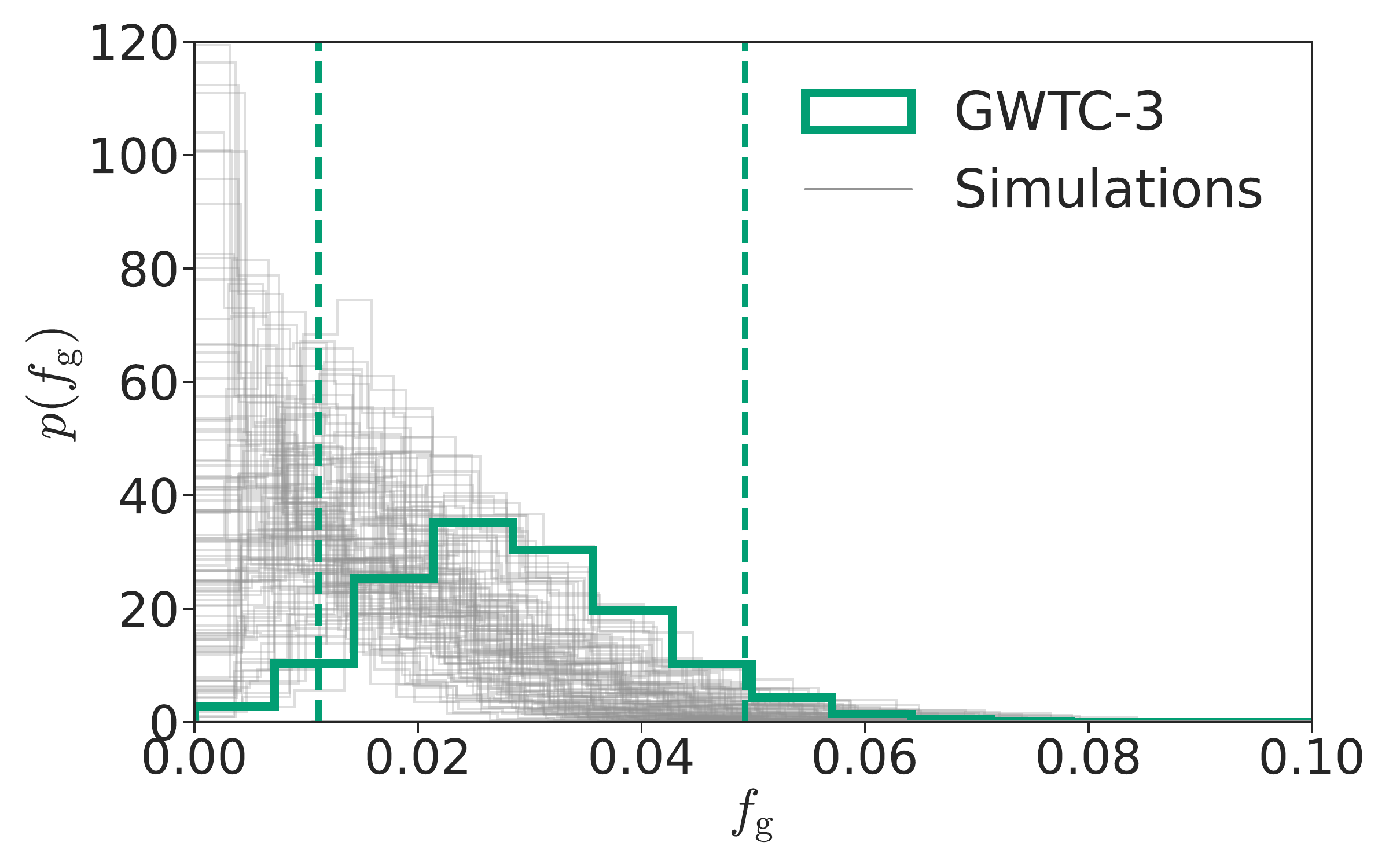}
\caption{\label{fig:histogram_bump} Marginalized posterior distributions (gray) for each of the 100 realizations of our GWTC-3-like simulations on the mixture model weight $f_{\rm{g}}$ for the Gaussian component in the \plplpeak model. In green, we show the measured posterior on $f_{\rm{g}}$ with GWTC-3 data.}
\end{center}
\end{figure}

To quantify the fraction of realizations which can reproduce our results due to clustering of events around $70~M_{\odot}$ when no such sub-population is present, in Fig.~\ref{fig:histogram_bump} we show the marginalized posterior distribution on $f_{\rm{g}}$ for both our simulations and GWTC-3. We find that for GWTC-3, $f_{\rm g}>0.01$ is at 96\% confidence. For our simulations, only 1 of the realizations have a similar significance of $f_{\rm{g}}>0.01$ at 96\% confidence.

\section{Does spin tell us anything?}\label{sec:spin}
Apart from masses and redshifts, BBH observations provide us with information about each black hole spin orientation and magnitude through the effective inspiral spin parameter $\chieff$ defined as,
\begin{equation}
    \chi_{\rm{eff}} = \frac{m_1\chi_{1,z} + m_2\chi_{2,z}}{m_1 + m_2}
\end{equation}
where $\chi_{1,z}$ and $\chi_{2,z}$ are the  dimensionless spin vector components (projected along the angular momentum axis of the binary), for the primary and secondary BH respectively. In general, $\chi_{\rm{eff}}$ is a measure for the alignment ($\chieff > 0$) or misalignment ($\chieff < 0$) of individual BH spins. 

From a population point of view, the distribution of $\chieff$ can provide hints as to the origin and formation channels of BBH mergers. A positively skewed $\chieff$ distribution is consistent with the majority of mergers having positively aligned spins, hinting at formation in the field \citep{Gerosa_2018,Zaldarriaga_2018,Belczynski_2020}, whereas a distribution which is symmetric about zero, hints at dynamically formed binaries due to the spin tilts being isotropically distributed at binary assembly \citep{Rodriguez_2015,Antonini_2016,Gerosa_2017}. 

Thus to probe the origin of the sub-populations discussed in Sec.~\ref{sec:Results} for the \plplpeak model, we consider extended models $p(m,\chi_{\rm{eff}}|\Lambda)$ which fit the population distribution of $\chi_{\rm{eff}}$ for each component in our mixture models. We model the $\chi_{\rm{eff}}$ distribution as a truncated Gaussian distribution in the range $\chieff \in [-1,1]$ for each sub-population, that is, we have three distinct $\chieff$ distributions with population parameters $\{\mu_{\chi,1},\sigma_{\chi,1},\mu_{\chi,2},\sigma_{\chi,2},\mu_{\chi,\rm{g}},\sigma_{\chi,\rm{g}} \}$ describing the location and width of the $\chieff$ distributions for the first and second power law, as well as the Gaussian bump, respectively.

\begin{figure}
\begin{center}
\includegraphics[width=0.48\textwidth]{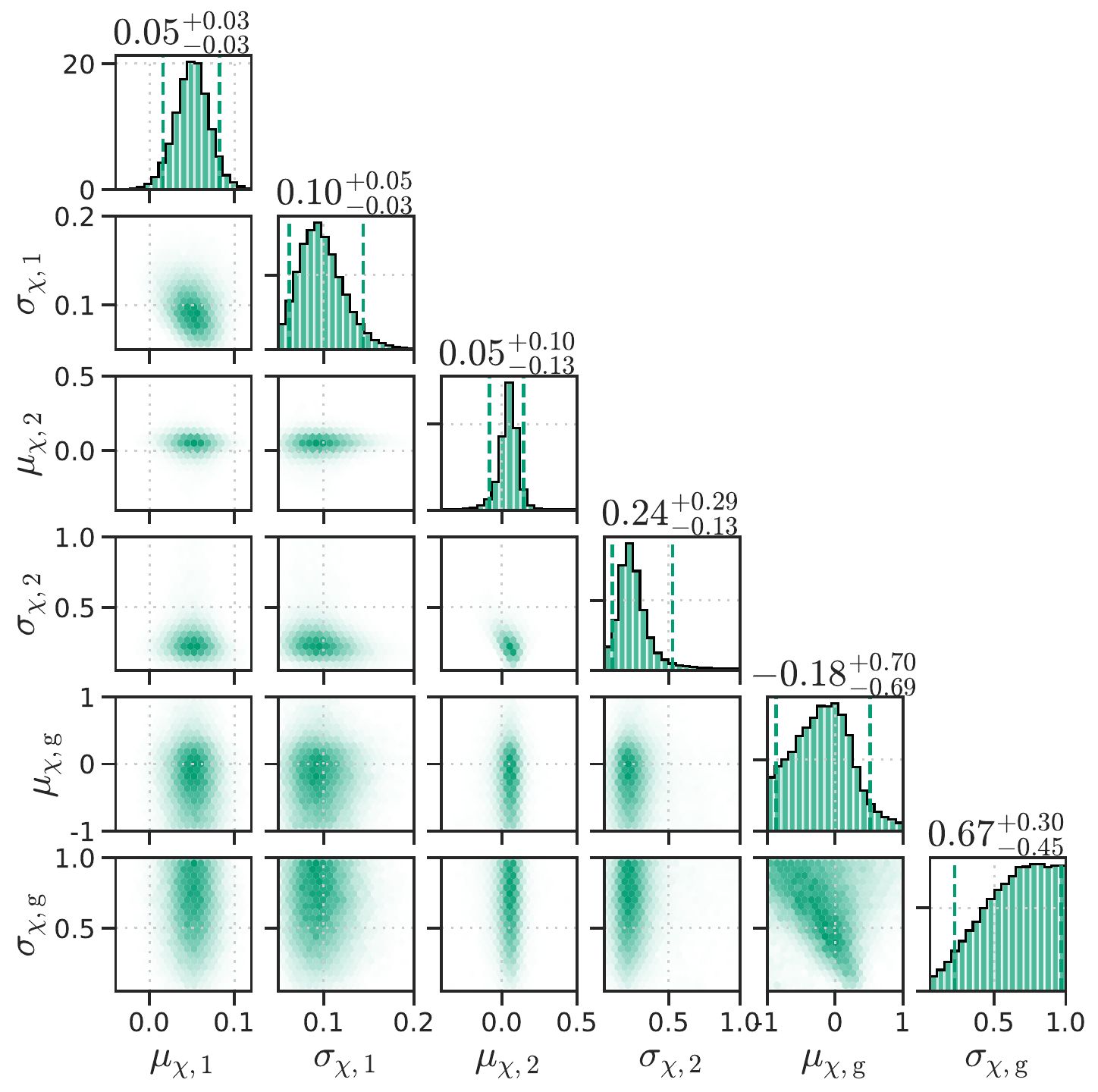}
\caption{\label{fig:chieff_corner} Marginalized posterior distribution on the locations and standard deviations for the Gaussian $\chieff$ population distributions for each mixture component of the \plplpeak model.}
\end{center}
\end{figure}

\begin{figure}
\begin{center}
\includegraphics[width=0.45\textwidth]{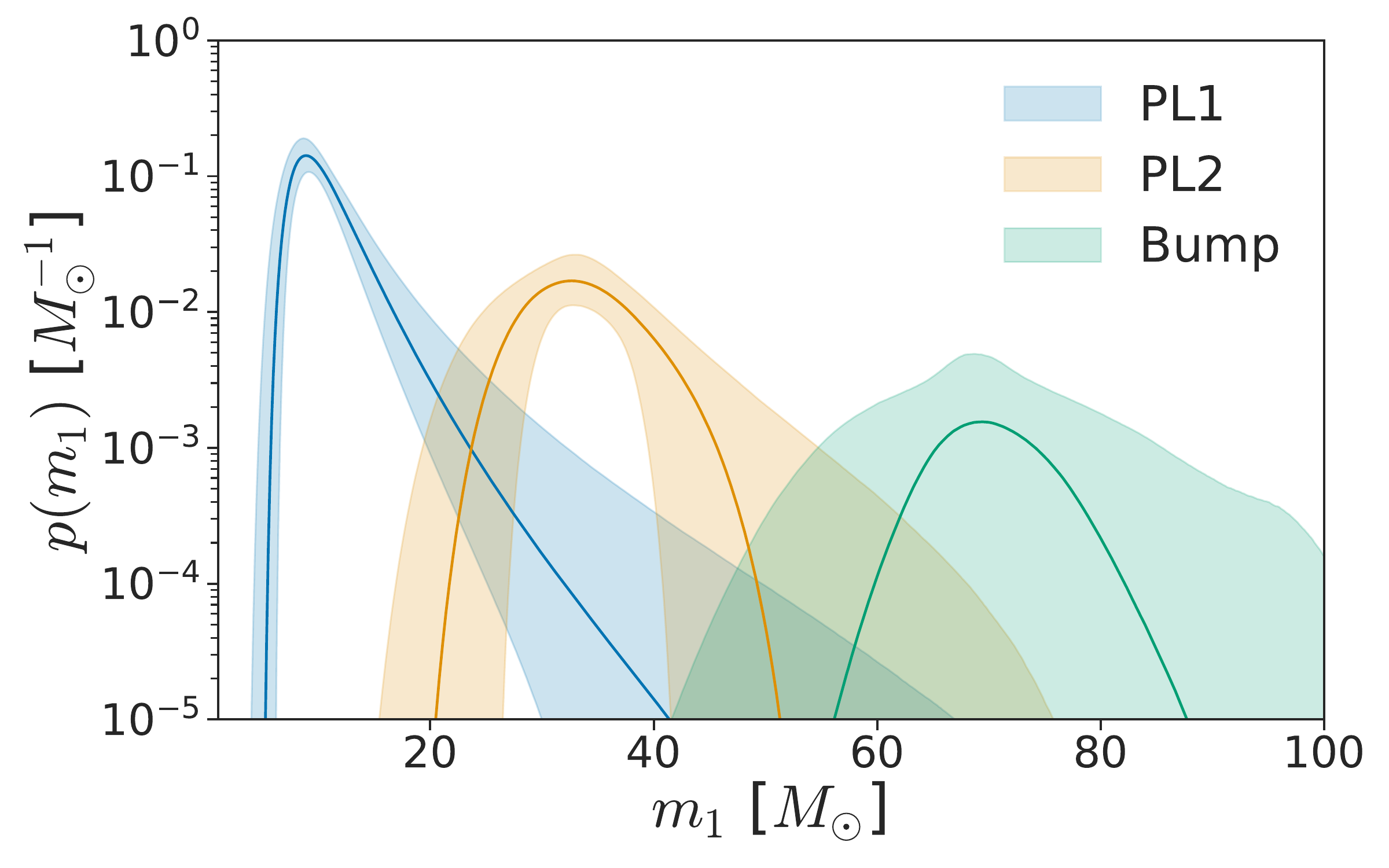}
\includegraphics[width=0.45\textwidth]{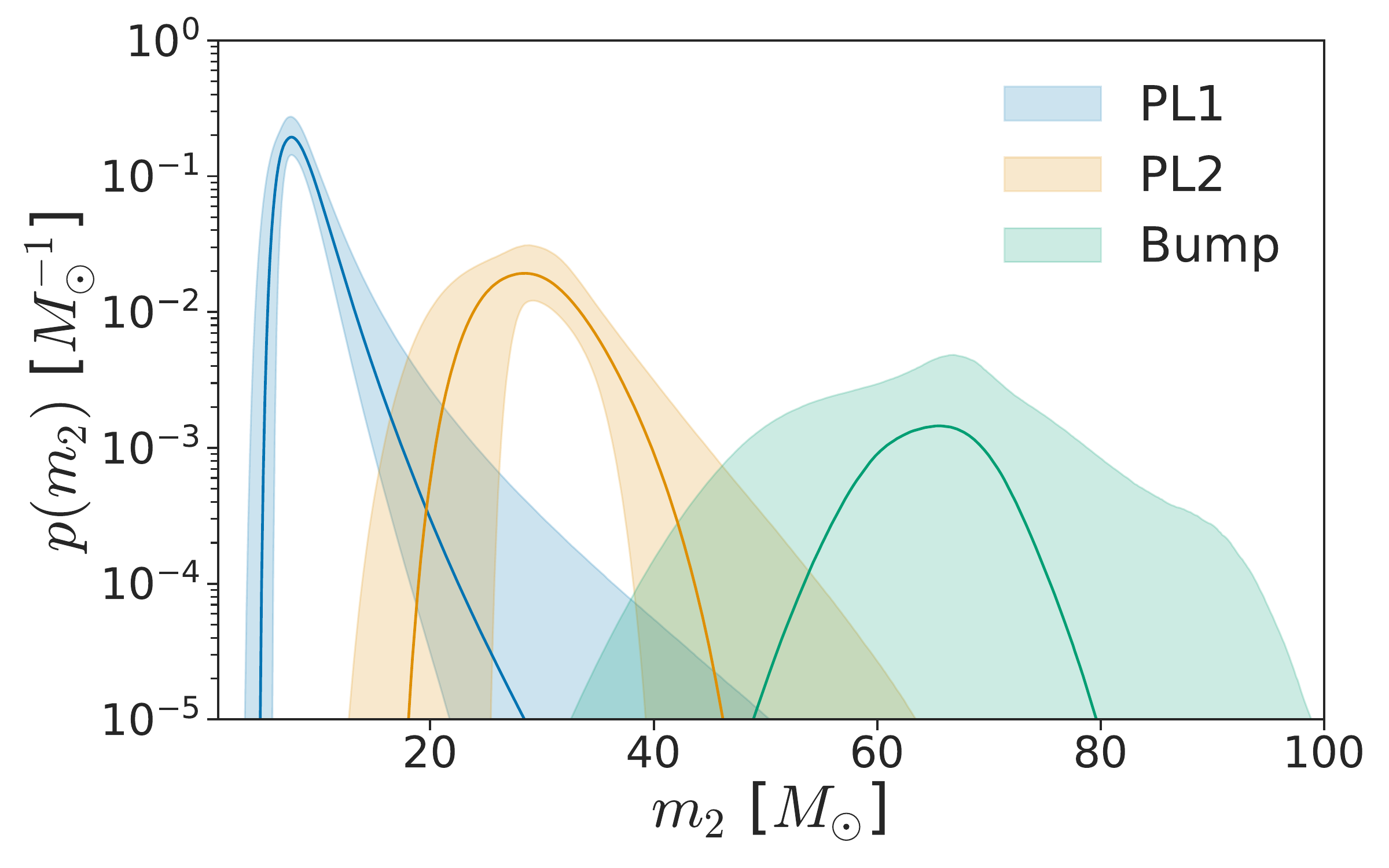}
\caption{\label{fig:mass_dists} Reconstructed $m_1$ and $m_2$ population distributions for each of the components in the \plplpeak model.}
\end{center}
\end{figure}

\begin{figure}
\begin{center}
\includegraphics[width=0.45\textwidth]{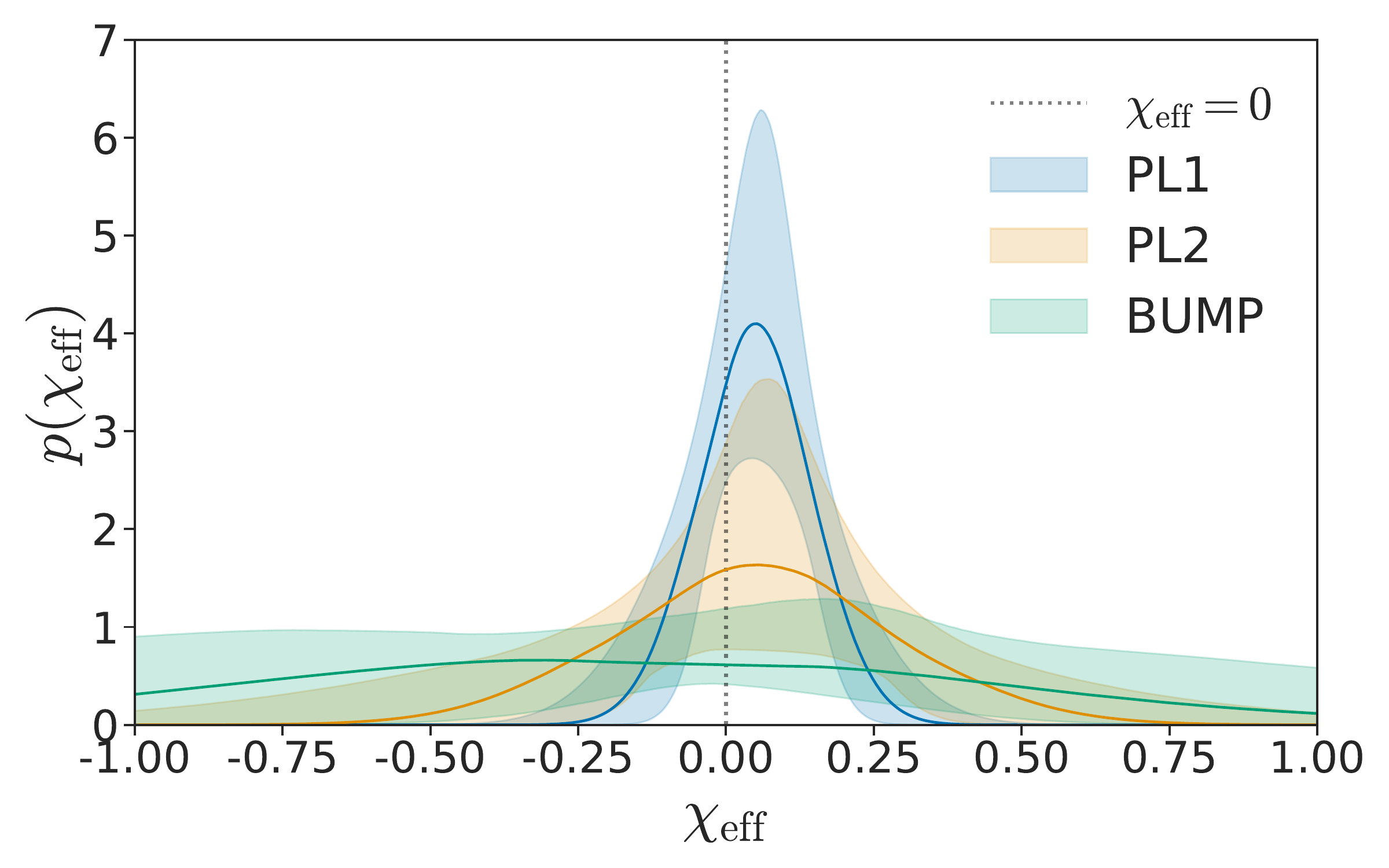}
\includegraphics[width=0.45\textwidth]{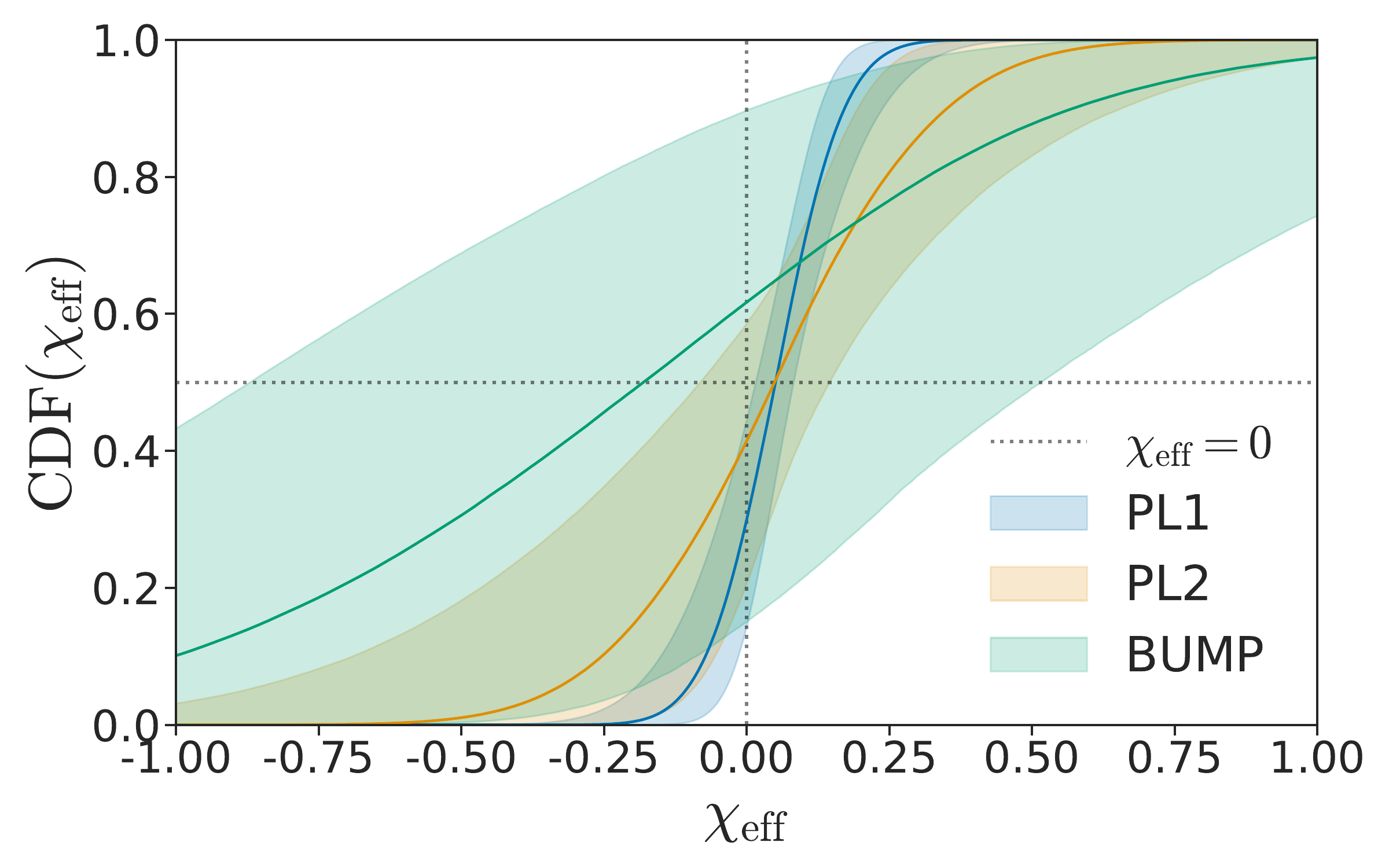}
\caption{\label{fig:chieff_dists} Reconstructed $\chieff$ population distributions for each of the components in the \plplpeak model. We also show the $\text{CDF}(\chieff)$ to more easily demonstrate the symmetry (about $\chieff=0$) for each sub-population.}
\end{center}
\end{figure}

In Fig.~\ref{fig:chieff_corner}, we show the posterior distribution on the $\chieff$ population distribution parameters. In Fig.~\ref{fig:mass_dists}, we show the reconstructed primary and secondary mass population distributions for each of the components for the \plplpeak model. Similarly, in Fig.~\ref{fig:chieff_dists} we show the reconstructed $\chieff$ population distribution for each of the components for the \plplpeak model. From our results, we find that $\mu_{\chi,1}>0$ at $>99.8$\% confidence for the first power law (PL1) component and to be constrained within the range $[0.02, 0.08]$ at 90\% confidence. Whereas for the second power law (PL2) and Gaussian Bump we find $\mu_{\chi,2}$ and $\mu_{\chi,\rm{g}}$ to be constrained to the within the ranges $[-0.08, 0.15]$ and $[-0.87, 0.52]$ respectively at 90\% confidence. We find this population to be broad but skewed towards negative $\chieff$. However, more data in this region of parameter space is needed to make definitive conclusions regarding the origin of BHs in this sub-population.

Consequently, when we look at the reconstructed component distributions in primary mass and $\chieff$, we see that PL1 is slightly skewed towards positive $\chieff$ and is well constrained within the range $[-0.2, 0.2]$. As for PL2, we find that it is consistent with a distribution symmetric about zero but has some preference for positive $\chieff$. This is not surprising, as the mass range for $m_1$ for PL2 is in the range $20-60~M_{\odot}$ and may include higher mass objects formed through dynamical interaction as well as some lower mass objects from field binaries. More interestingly, for the sub-population at $70~M_{\odot}$, we find a broad and symmetric distribution on $\chieff$. 

\section{Astrophysical Implications}\label{sec:astro}

\subsection{Low-mass $\sim 10-20~M_\odot$ population}
We start by discussing the first power law, PL1, which appears to be consistent with the expectations for isolated binaries~\citep{2022ApJ...931...17V}, and in agreement with findings from previous works searching for isolated binaries from this sub-population using GWTC-3 \citep{godfrey2024cosmiccousinsidentificationsubpopulation,li2024,ray2024searchingbinaryblackhole}. This population is likely to drop more steeply with mass than what observed in \citep{gwtc3_sensitivity} for the broader power law, with a maximum mass buried under PL2. Its effective spin distribution is likely (at 90\% CI) centered at values larger than zero, therefore supporting marginally aligned (with the binary angular momentum) spins and/or slowly spinning black holes. Aligned spins are typically predicted within the isolated binary formation channel which is in agreement with our observations. On this other hand, assuming this origin, the relatively low values may imply efficient angular momentum transport between the envelope and the core of the progenitor star leading to black holes that are slowly spinning \citep{bavera2020,Belczynski_2020}, and supernova kicks introducing misalignment between spins and binary angular momentum \citep{kalogera_1996}. Note that other formation channels are also expected to produce BBHs within this population, so we cannot assume that the entirety of this population (comprising $\sim 76\%$ of the BBH population according to the \plplpeak model) originates in the field.

\subsection{Intermediate-mass $\sim 20-40~M_\odot$ population}

Concerning the PL2 population, we are interpreting the $\sim30~M_\odot$ peak as the onset of a second BH population,  rather than a special feature on its own. The \plplpeak model favors a maximum mass at $\sim45-50~M_\odot$, while the \plpeak reaches $\sim100~M_\odot$. Of course in the latter case, the model needs to also fit for the high mass mergers with component masses at $\sim60~M_\odot$, so the second PL needs a higher maximum mass. Interestingly, the maximum mass for the \plplpeak is $m_{\rm{max},2} = 58^{+32}_{-14}~M_\odot$, and is consistent with the expectation for the onset of the pair--instability supernova (PISN) mass gap \citep{Farmer_2019,Farag:2022jcc}, in which case it is possible that the PL2 population also originates from isolated binaries, or is made of first generation dynamically-formed binaries. 

For PL2 from \plplpeak, we find a smoothing kernel of $\sim40~M_\odot$, revealing a less narrow feature than in the PLB model (found at $32.0^{+1.9}_{-2.0}~M_\odot$ with width $3.1^{+3.9}_{-1.8}~M_\odot$ in the \plpeak model) which may be more easily explained by popular formation channels, which typically do not predict such a narrow feature at $\sim30-35~M_\odot$.

Various formation channels may be able to reproduce a second $>20~M_\odot$ mass population in the mass distribution which is marginally better represented by a second power law rather than a Gaussian bump. Note that this is different from a broken power law as the amplitude of the second power law is free to vary, such that this second population does not necessarily need to be connected to the one peaking at $\sim 10~M_\odot$.
\cite{2022MNRAS.516.5737B,2022ApJ...931...17V,2023ApJ...948..105V,2023ApJ...955..127C} show that binary stellar population models can produce various mass distribution features, including a $\sim30-35~M_\odot$ peak. For instance, depending on whether the binary underwent a common envelope phase (CE), one could distinguish between an isolated binary CE subchannel and a stable Roche-lobe overflow (RLOF) subchannel \citep{2022ApJ...931...17V}, which may explain the emergence of two power laws below and above $\sim30~M_\odot$. Additionally, dense stellar clusters, including nuclear and globular clusters, can predict a sub-population peaking around this mass \citep{Mapelli_2022,2023ApJ...955..127C}.

The spin distribution is relatively inconclusive with respect to the origin of PL2. Although consistent with being centered at zero, it is still marginally skewed towards aligned spins - possibly because it originates from dynamical channels but is contaminated by objects which belong to the lower mass population, or because this sub-population also originates in the field. The larger width of the distribution compared to PL1 is compelling for a dynamical origin. Recent work~\citep{Biscoveanu_2022,2024PhRvD.109j3006H} finds a correlation between the width of the effective spin distribution and redshift. We argue that this may be related to what we observe here. If we allow for the different sub-populations to vary with redshift independently of each other, we find that at higher redshift the lower mass population rate decreases as also found by~\citep{rinaldi_2024}, while the PL2 rate increases. While such redshift evolution may not be physical due to the limited detector horizon especially at lower masses \citep{Heinzel_2024}, this finding may reconcile our and previous works' observations that the second sub-population drives the spreading of the spin width at higher redshift. 
However, note that uncertainties on spins for higher mass systems are typically larger than those from PL1, so it is challenging to draw definitive conclusions. Perhaps the most compelling evidence that PL2 may be composed of isolated binaries or dynamically formed first generation binaries is the fact that the PLPLB model PL2 has a maximum mass consistent with PISN mass gap. An involved study exploring the redshift and $\chieff$ dependence on the mass distribution is ongoing.

We note that \citep{godfrey2024cosmiccousinsidentificationsubpopulation} and \citep{li2024} also find evidence for 2 distinct sub-populations.  Specifically, \citep{godfrey2024cosmiccousinsidentificationsubpopulation} finds that the higher mass population which we identify here with PL2 shares characteristics with the likely isolated binary population peaking at $\sim 10~M_\odot$. Given all of the arguments and previous works outlined, we argue that PL1 is composed of isolated binaries, while PL2 may mostly arise from 1g+1g binaries from dynamical formation.

\subsection{High-mass $\sim 60-80~M_\odot$ population}

Although it may be tempting to identify the feature at $\sim70~M_\odot$ as the pulsational PISN (PPISN) build-up \citep{Talbot:2018cva}, the bump is located at a mass too high to be at the lower edge of the PISN mass gap, which is expected to be around $45~M_\odot$ and likely below $56~M_\odot$ considering reasonable variations in metallicity and uncertainties in the CO reaction rates \citep{Farmer_2019}. For the high mass bump, we find a broad distribution on $\chieff$. Although with large uncertainties given the current statistics, this population supports the presence of effective spins $<-0.3$, which are expected to exist for hierarchical mergers \citep{Baibhav_2020,Fishbach_2022}, but are hardly explained by isolated binary formation. Given these arguments, we argue that a plausible explanation for this feature is from hierarchical mergers of BHs from the $\sim30-35~M_\odot$ bump.  Because we find that the fraction of BBHs in the bump is 1-5\% at 90\% confidence, assuming these are from 2g BHs in dense star clusters, it is likely that these BHs are born spinning ($\chi_{\rm birth}>0$), but not enough to wash out this feature completely ($\chi_{\rm birth}<0.5$), according to the predictions of ~\citep{rodriguez19}. A high mass distribution similar to what found here is also predicted by~\citep{Kremer2020}, with the bump above $\sim 60~M_\odot$ originating from 2g hierarchical mergers and BHs born from stellar collision.

\section{Conclusions}\label{sec:conclusion}
In this paper we have explored alternative mass models for the BBH mass population. In particular, we construct a mixture model of power law and Gaussian distributions and fit the confident GWTC-3 BBH observations to date. We find that a mixture of two power law distributions is equally preferred by the GWTC-3 data compared to a power law plus Gaussian peak.  The first power law is likely  comprised by a large fraction of field binaries, which is consistent with previous findings, while we argue that the second power law includes a significant fraction of dynamically assembled first generation compact object binaries. However, our simulations show that a mixture of two power laws may model an underlying \plpeak population similarly well to what we find on the data for $\sim 7\%$ of our representations, and claim that more data is needed to robustly assess whether the \plpl model is indeed preferred over the \plplpeak one.

We also find that adding a Gaussian component above $40~M_\odot$ robustly finds a sub-population located around $65-70~M_\odot$. It is possible that by allowing a second power law to exist with a steeper slope than the first PL, one enables the detection of a sub-population that was otherwise buried under a single, flatter power law. This sub-population may arise from second generation hierarchical mergers, forming from black holes in the secondary peak of the distribution at $\sim 35~M_\odot$. Our simulations show that a similar feature cannot be recovered by chance coincidence when the underlying population follows a \plpeak. However, more data is required to confidently claim that the \plplpeak model is preferred over other models. We expect that the increasing number of BBH events detected through O4 and beyond will provide the observations required to determine whether the $70~M_\odot$ sub-population is actually present in the astrophysical BH mass distribution. If confirmed, such sub-population would a smoking gun about the dynamical origin of high-mass LVK BBHs, and may represent a new valuable feature to improve cosmological parameters constraints through spectral standard siren measurements \citep{Farr_2019,spectral_sirens}.

\section{Acknowledgements}
The authors would like to thank Anarya Ray, Katelyn Breivik, Caitlin Rose and Yeajin Kim for useful comments. IMH is supported by a McWilliams postdoctoral fellowship at Carnegie Mellon University. This material is based upon work supported by the National Aeronautics and Space Administration under Grant No. 22-LPS22-0025. This research has made use of data or software obtained from the Gravitational Wave Open Science Center (gwosc.org), a service of the LIGO Scientific Collaboration, the Virgo Collaboration, and KAGRA. This material is based upon work supported by NSF's LIGO Laboratory which is a major facility fully funded by the National Science Foundation, as well as the Science and Technology Facilities Council (STFC) of the United Kingdom, the Max-Planck-Society (MPS), and the State of Niedersachsen/Germany for support of the construction of Advanced LIGO and construction and operation of the GEO600 detector. Additional support for Advanced LIGO was provided by the Australian Research Council. Virgo is funded, through the European Gravitational Observatory (EGO), by the French Centre National de Recherche Scientifique (CNRS), the Italian Istituto Nazionale di Fisica Nucleare (INFN) and the Dutch Nikhef, with contributions by institutions from Belgium, Germany, Greece, Hungary, Ireland, Japan, Monaco, Poland, Portugal, Spain. KAGRA is supported by Ministry of Education, Culture, Sports, Science and Technology (MEXT), Japan Society for the Promotion of Science (JSPS) in Japan; National Research Foundation (NRF) and Ministry of Science and ICT (MSIT) in Korea; Academia Sinica (AS) and National Science and Technology Council (NSTC) in Taiwan.

\bibliography{references}{}

\begin{thebibliography}{71}
\expandafter\ifx\csname natexlab\endcsname\relax\def\natexlab#1{#1}\fi
\expandafter\ifx\csname bibnamefont\endcsname\relax
  \def\bibnamefont#1{#1}\fi
\expandafter\ifx\csname bibfnamefont\endcsname\relax
  \def\bibfnamefont#1{#1}\fi
\expandafter\ifx\csname citenamefont\endcsname\relax
  \def\citenamefont#1{#1}\fi
\expandafter\ifx\csname url\endcsname\relax
  \def\url#1{\texttt{#1}}\fi
\expandafter\ifx\csname urlprefix\endcsname\relax\def\urlprefix{URL }\fi
\providecommand{\bibinfo}[2]{#2}
\providecommand{\eprint}[2][]{\url{#2}}

\bibitem[{\citenamefont{Abbott et~al.}(2016{\natexlab{a}})}]{LIGOScientific:2016aoc}
\bibinfo{author}{\bibfnamefont{B.~P.} \bibnamefont{Abbott}} \bibnamefont{et~al.} (\bibinfo{collaboration}{LIGO Scientific, Virgo}), \bibinfo{journal}{Phys. Rev. Lett.} \textbf{\bibinfo{volume}{116}}, \bibinfo{pages}{061102} (\bibinfo{year}{2016}{\natexlab{a}}), \eprint{1602.03837}.

\bibitem[{\citenamefont{Abbott et~al.}(2023{\natexlab{a}})}]{gwtc3_data}
\bibinfo{author}{\bibfnamefont{R.}~\bibnamefont{Abbott}} \bibnamefont{et~al.}, \emph{\bibinfo{title}{{GWTC-3: Compact Binary Coalescences Observed by LIGO and Virgo During the Second Part of the Third Observing Run — Parameter estimation data release}}} (\bibinfo{year}{2023}{\natexlab{a}}), \urlprefix\url{https://doi.org/10.5281/zenodo.8177023}.

\bibitem[{\citenamefont{Rodriguez et~al.}(2015)\citenamefont{Rodriguez, Morscher, Pattabiraman, Chatterjee, Haster, and Rasio}}]{Rodriguez_2015}
\bibinfo{author}{\bibfnamefont{C.~L.} \bibnamefont{Rodriguez}}, \bibinfo{author}{\bibfnamefont{M.}~\bibnamefont{Morscher}}, \bibinfo{author}{\bibfnamefont{B.}~\bibnamefont{Pattabiraman}}, \bibinfo{author}{\bibfnamefont{S.}~\bibnamefont{Chatterjee}}, \bibinfo{author}{\bibfnamefont{C.-J.} \bibnamefont{Haster}}, \bibnamefont{and} \bibinfo{author}{\bibfnamefont{F.~A.} \bibnamefont{Rasio}}, \bibinfo{journal}{Physical Review Letters} \textbf{\bibinfo{volume}{115}} (\bibinfo{year}{2015}), ISSN \bibinfo{issn}{1079-7114}, \urlprefix\url{http://dx.doi.org/10.1103/PhysRevLett.115.051101}.

\bibitem[{\citenamefont{Rodriguez et~al.}(2019)\citenamefont{Rodriguez, Zevin, Amaro-Seoane, Chatterjee, Kremer, Rasio, and Ye}}]{rodriguez19}
\bibinfo{author}{\bibfnamefont{C.~L.} \bibnamefont{Rodriguez}}, \bibinfo{author}{\bibfnamefont{M.}~\bibnamefont{Zevin}}, \bibinfo{author}{\bibfnamefont{P.}~\bibnamefont{Amaro-Seoane}}, \bibinfo{author}{\bibfnamefont{S.}~\bibnamefont{Chatterjee}}, \bibinfo{author}{\bibfnamefont{K.}~\bibnamefont{Kremer}}, \bibinfo{author}{\bibfnamefont{F.~A.} \bibnamefont{Rasio}}, \bibnamefont{and} \bibinfo{author}{\bibfnamefont{C.~S.} \bibnamefont{Ye}}, \bibinfo{journal}{Phys. Rev. D} \textbf{\bibinfo{volume}{100}}, \bibinfo{pages}{043027} (\bibinfo{year}{2019}), \urlprefix\url{https://link.aps.org/doi/10.1103/PhysRevD.100.043027}.

\bibitem[{\citenamefont{McKernan et~al.}(2012)\citenamefont{McKernan, Ford, Lyra, and Perets}}]{2012MNRAS.425..460M}
\bibinfo{author}{\bibfnamefont{B.}~\bibnamefont{McKernan}}, \bibinfo{author}{\bibfnamefont{K.~E.~S.} \bibnamefont{Ford}}, \bibinfo{author}{\bibfnamefont{W.}~\bibnamefont{Lyra}}, \bibnamefont{and} \bibinfo{author}{\bibfnamefont{H.~B.} \bibnamefont{Perets}}, \bibinfo{journal}{Monthly Notices of the Royal Astronomical Society} \textbf{\bibinfo{volume}{425}}, \bibinfo{pages}{460–469} (\bibinfo{year}{2012}), ISSN \bibinfo{issn}{0035-8711}, \urlprefix\url{http://dx.doi.org/10.1111/j.1365-2966.2012.21486.x}.

\bibitem[{\citenamefont{Antonini and Rasio}(2016)}]{Antonini_2016}
\bibinfo{author}{\bibfnamefont{F.}~\bibnamefont{Antonini}} \bibnamefont{and} \bibinfo{author}{\bibfnamefont{F.~A.} \bibnamefont{Rasio}}, \bibinfo{journal}{The Astrophysical Journal} \textbf{\bibinfo{volume}{831}}, \bibinfo{pages}{187} (\bibinfo{year}{2016}), ISSN \bibinfo{issn}{1538-4357}, \urlprefix\url{http://dx.doi.org/10.3847/0004-637X/831/2/187}.

\bibitem[{\citenamefont{McKernan et~al.}(2018)\citenamefont{McKernan, Saavik~Ford, Bellovary, Leigh, Haiman, Kocsis, Lyra, Mac~Low, Metzger, O’Dowd et~al.}}]{2018ApJ...866...66M}
\bibinfo{author}{\bibfnamefont{B.}~\bibnamefont{McKernan}}, \bibinfo{author}{\bibfnamefont{K.~E.} \bibnamefont{Saavik~Ford}}, \bibinfo{author}{\bibfnamefont{J.}~\bibnamefont{Bellovary}}, \bibinfo{author}{\bibfnamefont{N.~W.~C.} \bibnamefont{Leigh}}, \bibinfo{author}{\bibfnamefont{Z.}~\bibnamefont{Haiman}}, \bibinfo{author}{\bibfnamefont{B.}~\bibnamefont{Kocsis}}, \bibinfo{author}{\bibfnamefont{W.}~\bibnamefont{Lyra}}, \bibinfo{author}{\bibfnamefont{M.-M.} \bibnamefont{Mac~Low}}, \bibinfo{author}{\bibfnamefont{B.}~\bibnamefont{Metzger}}, \bibinfo{author}{\bibfnamefont{M.}~\bibnamefont{O’Dowd}}, \bibnamefont{et~al.}, \bibinfo{journal}{The Astrophysical Journal} \textbf{\bibinfo{volume}{866}}, \bibinfo{pages}{66} (\bibinfo{year}{2018}), ISSN \bibinfo{issn}{1538-4357}, \urlprefix\url{http://dx.doi.org/10.3847/1538-4357/aadae5}.

\bibitem[{\citenamefont{{Conselice} et~al.}(2020)\citenamefont{{Conselice}, {Bhatawdekar}, {Palmese}, and {Hartley}}}]{Conselice_20}
\bibinfo{author}{\bibfnamefont{C.~J.} \bibnamefont{{Conselice}}}, \bibinfo{author}{\bibfnamefont{R.}~\bibnamefont{{Bhatawdekar}}}, \bibinfo{author}{\bibfnamefont{A.}~\bibnamefont{{Palmese}}}, \bibnamefont{and} \bibinfo{author}{\bibfnamefont{W.~G.} \bibnamefont{{Hartley}}}, \bibinfo{journal}{The Astrophysical Journal} \textbf{\bibinfo{volume}{890}}, \bibinfo{eid}{arXiv:1907.05361} (\bibinfo{year}{2020}), \eprint{1907.05361}.

\bibitem[{\citenamefont{Palmese and Conselice}(2021)}]{palmese_conselice}
\bibinfo{author}{\bibfnamefont{A.}~\bibnamefont{Palmese}} \bibnamefont{and} \bibinfo{author}{\bibfnamefont{C.~J.} \bibnamefont{Conselice}}, \bibinfo{journal}{Physical Review Letters} \textbf{\bibinfo{volume}{126}} (\bibinfo{year}{2021}), ISSN \bibinfo{issn}{1079-7114}, \urlprefix\url{http://dx.doi.org/10.1103/PhysRevLett.126.181103}.

\bibitem[{\citenamefont{{Kinugawa} et~al.}(2014)\citenamefont{{Kinugawa}, {Inayoshi}, {Hotokezaka}, {Nakauchi}, and {Nakamura}}}]{Kinugawa2014MNRAS.442.2963K}
\bibinfo{author}{\bibfnamefont{T.}~\bibnamefont{{Kinugawa}}}, \bibinfo{author}{\bibfnamefont{K.}~\bibnamefont{{Inayoshi}}}, \bibinfo{author}{\bibfnamefont{K.}~\bibnamefont{{Hotokezaka}}}, \bibinfo{author}{\bibfnamefont{D.}~\bibnamefont{{Nakauchi}}}, \bibnamefont{and} \bibinfo{author}{\bibfnamefont{T.}~\bibnamefont{{Nakamura}}}, \bibinfo{journal}{Monthly Notices of the Royal Astronomical Society} \textbf{\bibinfo{volume}{442}}, \bibinfo{pages}{2963} (\bibinfo{year}{2014}), \eprint{1402.6672}.

\bibitem[{\citenamefont{Hartwig et~al.}(2016)\citenamefont{Hartwig, Volonteri, Bromm, Klessen, Barausse, Magg, and Stacy}}]{Hartwig_2016}
\bibinfo{author}{\bibfnamefont{T.}~\bibnamefont{Hartwig}}, \bibinfo{author}{\bibfnamefont{M.}~\bibnamefont{Volonteri}}, \bibinfo{author}{\bibfnamefont{V.}~\bibnamefont{Bromm}}, \bibinfo{author}{\bibfnamefont{R.~S.} \bibnamefont{Klessen}}, \bibinfo{author}{\bibfnamefont{E.}~\bibnamefont{Barausse}}, \bibinfo{author}{\bibfnamefont{M.}~\bibnamefont{Magg}}, \bibnamefont{and} \bibinfo{author}{\bibfnamefont{A.}~\bibnamefont{Stacy}}, \bibinfo{journal}{Monthly Notices of the Royal Astronomical Society: Letters} \textbf{\bibinfo{volume}{460}}, \bibinfo{pages}{L74–L78} (\bibinfo{year}{2016}), ISSN \bibinfo{issn}{1745-3933}, \urlprefix\url{http://dx.doi.org/10.1093/mnrasl/slw074}.

\bibitem[{\citenamefont{Carr and Hawking}(1974)}]{Carr:1974nx}
\bibinfo{author}{\bibfnamefont{B.~J.} \bibnamefont{Carr}} \bibnamefont{and} \bibinfo{author}{\bibfnamefont{S.~W.} \bibnamefont{Hawking}}, \bibinfo{journal}{Mon. Not. Roy. Astron. Soc.} \textbf{\bibinfo{volume}{168}}, \bibinfo{pages}{399} (\bibinfo{year}{1974}).

\bibitem[{\citenamefont{Clesse and García-Bellido}(2017)}]{Clesse:2016ajp}
\bibinfo{author}{\bibfnamefont{S.}~\bibnamefont{Clesse}} \bibnamefont{and} \bibinfo{author}{\bibfnamefont{J.}~\bibnamefont{García-Bellido}}, \bibinfo{journal}{Phys. Dark Univ.} \textbf{\bibinfo{volume}{18}}, \bibinfo{pages}{105} (\bibinfo{year}{2017}), \eprint{1610.08479}.

\bibitem[{\citenamefont{Tsai et~al.}(2021)\citenamefont{Tsai, Palmese, Profumo, and Jeltema}}]{Tsai_2021}
\bibinfo{author}{\bibfnamefont{Y.-D.} \bibnamefont{Tsai}}, \bibinfo{author}{\bibfnamefont{A.}~\bibnamefont{Palmese}}, \bibinfo{author}{\bibfnamefont{S.}~\bibnamefont{Profumo}}, \bibnamefont{and} \bibinfo{author}{\bibfnamefont{T.}~\bibnamefont{Jeltema}}, \bibinfo{journal}{Journal of Cosmology and Astroparticle Physics} \textbf{\bibinfo{volume}{2021}}, \bibinfo{pages}{019} (\bibinfo{year}{2021}), ISSN \bibinfo{issn}{1475-7516}, \urlprefix\url{http://dx.doi.org/10.1088/1475-7516/2021/10/019}.

\bibitem[{\citenamefont{Mapelli}(2021)}]{Mapelli_2021}
\bibinfo{author}{\bibfnamefont{M.}~\bibnamefont{Mapelli}}, \emph{\bibinfo{title}{Formation Channels of Single and Binary Stellar-Mass Black Holes}} (\bibinfo{publisher}{Springer Singapore}, \bibinfo{year}{2021}), p. \bibinfo{pages}{1–65}, ISBN \bibinfo{isbn}{9789811547027}, \urlprefix\url{http://dx.doi.org/10.1007/978-981-15-4702-7_16-1}.

\bibitem[{\citenamefont{Mandel and Farmer}(2022)}]{Mandel_2022}
\bibinfo{author}{\bibfnamefont{I.}~\bibnamefont{Mandel}} \bibnamefont{and} \bibinfo{author}{\bibfnamefont{A.}~\bibnamefont{Farmer}}, \bibinfo{journal}{Physics Reports} \textbf{\bibinfo{volume}{955}}, \bibinfo{pages}{1–24} (\bibinfo{year}{2022}), ISSN \bibinfo{issn}{0370-1573}, \urlprefix\url{http://dx.doi.org/10.1016/j.physrep.2022.01.003}.

\bibitem[{\citenamefont{Zevin et~al.}(2021)\citenamefont{Zevin, Bavera, Berry, Kalogera, Fragos, Marchant, Rodriguez, Antonini, Holz, and Pankow}}]{Zevin_2021}
\bibinfo{author}{\bibfnamefont{M.}~\bibnamefont{Zevin}}, \bibinfo{author}{\bibfnamefont{S.~S.} \bibnamefont{Bavera}}, \bibinfo{author}{\bibfnamefont{C.~P.~L.} \bibnamefont{Berry}}, \bibinfo{author}{\bibfnamefont{V.}~\bibnamefont{Kalogera}}, \bibinfo{author}{\bibfnamefont{T.}~\bibnamefont{Fragos}}, \bibinfo{author}{\bibfnamefont{P.}~\bibnamefont{Marchant}}, \bibinfo{author}{\bibfnamefont{C.~L.} \bibnamefont{Rodriguez}}, \bibinfo{author}{\bibfnamefont{F.}~\bibnamefont{Antonini}}, \bibinfo{author}{\bibfnamefont{D.~E.} \bibnamefont{Holz}}, \bibnamefont{and} \bibinfo{author}{\bibfnamefont{C.}~\bibnamefont{Pankow}}, \bibinfo{journal}{The Astrophysical Journal} \textbf{\bibinfo{volume}{910}}, \bibinfo{pages}{152} (\bibinfo{year}{2021}), ISSN \bibinfo{issn}{1538-4357}, \urlprefix\url{http://dx.doi.org/10.3847/1538-4357/abe40e}.

\bibitem[{\citenamefont{{Cheng} et~al.}(2023)\citenamefont{{Cheng}, {Zevin}, and {Vitale}}}]{2023ApJ...955..127C}
\bibinfo{author}{\bibfnamefont{A.~Q.} \bibnamefont{{Cheng}}}, \bibinfo{author}{\bibfnamefont{M.}~\bibnamefont{{Zevin}}}, \bibnamefont{and} \bibinfo{author}{\bibfnamefont{S.}~\bibnamefont{{Vitale}}}, \bibinfo{journal}{The Astrophysical Journal} \textbf{\bibinfo{volume}{955}}, \bibinfo{eid}{127} (\bibinfo{year}{2023}), \eprint{2307.03129}.

\bibitem[{\citenamefont{Ray et~al.}(2024)\citenamefont{Ray, Hernandez, Breivik, and Creighton}}]{ray2024searchingbinaryblackhole}
\bibinfo{author}{\bibfnamefont{A.}~\bibnamefont{Ray}}, \bibinfo{author}{\bibfnamefont{I.~M.} \bibnamefont{Hernandez}}, \bibinfo{author}{\bibfnamefont{K.}~\bibnamefont{Breivik}}, \bibnamefont{and} \bibinfo{author}{\bibfnamefont{J.}~\bibnamefont{Creighton}}, \emph{\bibinfo{title}{Searching for binary black hole sub-populations in gravitational wave data using binned gaussian processes}} (\bibinfo{year}{2024}), \eprint{2404.03166}, \urlprefix\url{https://arxiv.org/abs/2404.03166}.

\bibitem[{\citenamefont{Abbott et~al.}(2021{\natexlab{a}})}]{LIGOScientific:2020kqk}
\bibinfo{author}{\bibfnamefont{R.}~\bibnamefont{Abbott}} \bibnamefont{et~al.} (\bibinfo{collaboration}{LIGO Scientific, Virgo}), \bibinfo{journal}{Astrophys. J. Lett.} \textbf{\bibinfo{volume}{913}}, \bibinfo{pages}{L7} (\bibinfo{year}{2021}{\natexlab{a}}), \eprint{2010.14533}.

\bibitem[{\citenamefont{Abbott et~al.}(2021{\natexlab{b}})}]{gwtc3_sensitivity}
\bibinfo{author}{\bibfnamefont{R.}~\bibnamefont{Abbott}} \bibnamefont{et~al.}, \emph{\bibinfo{title}{{GWTC-3: Compact Binary Coalescences Observed by LIGO and Virgo During the Second Part of the Third Observing Run — O3 search sensitivity estimates}}} (\bibinfo{year}{2021}{\natexlab{b}}), \urlprefix\url{https://doi.org/10.5281/zenodo.5546676}.

\bibitem[{\citenamefont{Baxter et~al.}(2021)\citenamefont{Baxter, Croon, McDermott, and Sakstein}}]{Baxter_2021}
\bibinfo{author}{\bibfnamefont{E.~J.} \bibnamefont{Baxter}}, \bibinfo{author}{\bibfnamefont{D.}~\bibnamefont{Croon}}, \bibinfo{author}{\bibfnamefont{S.~D.} \bibnamefont{McDermott}}, \bibnamefont{and} \bibinfo{author}{\bibfnamefont{J.}~\bibnamefont{Sakstein}}, \bibinfo{journal}{The Astrophysical Journal Letters} \textbf{\bibinfo{volume}{916}}, \bibinfo{pages}{L16} (\bibinfo{year}{2021}), ISSN \bibinfo{issn}{2041-8213}, \urlprefix\url{http://dx.doi.org/10.3847/2041-8213/ac11fc}.

\bibitem[{\citenamefont{Mandel et~al.}(2016)\citenamefont{Mandel, Farr, Colonna, Stevenson, Tiňo, and Veitch}}]{Mandel_2016}
\bibinfo{author}{\bibfnamefont{I.}~\bibnamefont{Mandel}}, \bibinfo{author}{\bibfnamefont{W.~M.} \bibnamefont{Farr}}, \bibinfo{author}{\bibfnamefont{A.}~\bibnamefont{Colonna}}, \bibinfo{author}{\bibfnamefont{S.}~\bibnamefont{Stevenson}}, \bibinfo{author}{\bibfnamefont{P.}~\bibnamefont{Tiňo}}, \bibnamefont{and} \bibinfo{author}{\bibfnamefont{J.}~\bibnamefont{Veitch}}, \bibinfo{journal}{Monthly Notices of the Royal Astronomical Society} \textbf{\bibinfo{volume}{465}}, \bibinfo{pages}{3254–3260} (\bibinfo{year}{2016}), ISSN \bibinfo{issn}{1365-2966}, \urlprefix\url{http://dx.doi.org/10.1093/mnras/stw2883}.

\bibitem[{\citenamefont{Edelman et~al.}(2023)\citenamefont{Edelman, Farr, and Doctor}}]{Edelman:2022ydv}
\bibinfo{author}{\bibfnamefont{B.}~\bibnamefont{Edelman}}, \bibinfo{author}{\bibfnamefont{B.}~\bibnamefont{Farr}}, \bibnamefont{and} \bibinfo{author}{\bibfnamefont{Z.}~\bibnamefont{Doctor}}, \bibinfo{journal}{Astrophys. J.} \textbf{\bibinfo{volume}{946}}, \bibinfo{pages}{16} (\bibinfo{year}{2023}), \eprint{2210.12834}.

\bibitem[{\citenamefont{Maga\~na Hernandez and Ray}(2024)}]{MaganaHernandez:2024uty}
\bibinfo{author}{\bibfnamefont{I.}~\bibnamefont{Maga\~na Hernandez}} \bibnamefont{and} \bibinfo{author}{\bibfnamefont{A.}~\bibnamefont{Ray}} (\bibinfo{year}{2024}), \eprint{2404.02522}.

\bibitem[{\citenamefont{Farah et~al.}(2024{\natexlab{a}})\citenamefont{Farah, Callister, Ezquiaga, Zevin, and Holz}}]{Farah:2024xub}
\bibinfo{author}{\bibfnamefont{A.~M.} \bibnamefont{Farah}}, \bibinfo{author}{\bibfnamefont{T.~A.} \bibnamefont{Callister}}, \bibinfo{author}{\bibfnamefont{J.~M.} \bibnamefont{Ezquiaga}}, \bibinfo{author}{\bibfnamefont{M.}~\bibnamefont{Zevin}}, \bibnamefont{and} \bibinfo{author}{\bibfnamefont{D.~E.} \bibnamefont{Holz}} (\bibinfo{year}{2024}{\natexlab{a}}), \eprint{2404.02210}.

\bibitem[{\citenamefont{Briel et~al.}(2023)\citenamefont{Briel, Stevance, and Eldridge}}]{Briel:2022cfl}
\bibinfo{author}{\bibfnamefont{M.~M.} \bibnamefont{Briel}}, \bibinfo{author}{\bibfnamefont{H.~F.} \bibnamefont{Stevance}}, \bibnamefont{and} \bibinfo{author}{\bibfnamefont{J.~J.} \bibnamefont{Eldridge}}, \bibinfo{journal}{Mon. Not. Roy. Astron. Soc.} \textbf{\bibinfo{volume}{520}}, \bibinfo{pages}{5724} (\bibinfo{year}{2023}), \eprint{2206.13842}.

\bibitem[{\citenamefont{Golomb et~al.}(2024)\citenamefont{Golomb, Isi, and Farr}}]{Golomg_2023}
\bibinfo{author}{\bibfnamefont{J.}~\bibnamefont{Golomb}}, \bibinfo{author}{\bibfnamefont{M.}~\bibnamefont{Isi}}, \bibnamefont{and} \bibinfo{author}{\bibfnamefont{W.~M.} \bibnamefont{Farr}}, \bibinfo{journal}{The Astrophysical Journal} \textbf{\bibinfo{volume}{976}}, \bibinfo{pages}{121} (\bibinfo{year}{2024}), ISSN \bibinfo{issn}{1538-4357}, \urlprefix\url{http://dx.doi.org/10.3847/1538-4357/ad8572}.

\bibitem[{\citenamefont{Hendriks et~al.}(2023)\citenamefont{Hendriks, van Son, Renzo, Izzard, and Farmer}}]{Hendriks:2023yrw}
\bibinfo{author}{\bibfnamefont{D.~D.} \bibnamefont{Hendriks}}, \bibinfo{author}{\bibfnamefont{L.~A.~C.} \bibnamefont{van Son}}, \bibinfo{author}{\bibfnamefont{M.}~\bibnamefont{Renzo}}, \bibinfo{author}{\bibfnamefont{R.~G.} \bibnamefont{Izzard}}, \bibnamefont{and} \bibinfo{author}{\bibfnamefont{R.}~\bibnamefont{Farmer}}, \bibinfo{journal}{Mon. Not. Roy. Astron. Soc.} \textbf{\bibinfo{volume}{526}}, \bibinfo{pages}{4130} (\bibinfo{year}{2023}), \eprint{2309.09339}.

\bibitem[{\citenamefont{Gerosa and Fishbach}(2021)}]{Gerosa:2021mno}
\bibinfo{author}{\bibfnamefont{D.}~\bibnamefont{Gerosa}} \bibnamefont{and} \bibinfo{author}{\bibfnamefont{M.}~\bibnamefont{Fishbach}}, \bibinfo{journal}{Nature Astron.} \textbf{\bibinfo{volume}{5}}, \bibinfo{pages}{749} (\bibinfo{year}{2021}), \eprint{2105.03439}.

\bibitem[{\citenamefont{Abbott et~al.}(2020{\natexlab{a}})}]{LIGOScientific:2020ufj}
\bibinfo{author}{\bibfnamefont{R.}~\bibnamefont{Abbott}} \bibnamefont{et~al.} (\bibinfo{collaboration}{LIGO Scientific, Virgo}), \bibinfo{journal}{Astrophys. J. Lett.} \textbf{\bibinfo{volume}{900}}, \bibinfo{pages}{L13} (\bibinfo{year}{2020}{\natexlab{a}}), \eprint{2009.01190}.

\bibitem[{\citenamefont{Kimball et~al.}(2020)\citenamefont{Kimball, Talbot, L.~Berry, Carney, Zevin, Thrane, and Kalogera}}]{Kimball_2020}
\bibinfo{author}{\bibfnamefont{C.}~\bibnamefont{Kimball}}, \bibinfo{author}{\bibfnamefont{C.}~\bibnamefont{Talbot}}, \bibinfo{author}{\bibfnamefont{C.~P.} \bibnamefont{L.~Berry}}, \bibinfo{author}{\bibfnamefont{M.}~\bibnamefont{Carney}}, \bibinfo{author}{\bibfnamefont{M.}~\bibnamefont{Zevin}}, \bibinfo{author}{\bibfnamefont{E.}~\bibnamefont{Thrane}}, \bibnamefont{and} \bibinfo{author}{\bibfnamefont{V.}~\bibnamefont{Kalogera}}, \bibinfo{journal}{The Astrophysical Journal} \textbf{\bibinfo{volume}{900}}, \bibinfo{pages}{177} (\bibinfo{year}{2020}), ISSN \bibinfo{issn}{1538-4357}, \urlprefix\url{http://dx.doi.org/10.3847/1538-4357/aba518}.

\bibitem[{\citenamefont{Kimball et~al.}(2021)}]{Kimball:2020qyd}
\bibinfo{author}{\bibfnamefont{C.}~\bibnamefont{Kimball}} \bibnamefont{et~al.}, \bibinfo{journal}{Astrophys. J. Lett.} \textbf{\bibinfo{volume}{915}}, \bibinfo{pages}{L35} (\bibinfo{year}{2021}), \eprint{2011.05332}.

\bibitem[{\citenamefont{Mandel et~al.}(2019)\citenamefont{Mandel, Farr, and Gair}}]{Mandel_2019}
\bibinfo{author}{\bibfnamefont{I.}~\bibnamefont{Mandel}}, \bibinfo{author}{\bibfnamefont{W.~M.} \bibnamefont{Farr}}, \bibnamefont{and} \bibinfo{author}{\bibfnamefont{J.~R.} \bibnamefont{Gair}}, \bibinfo{journal}{Monthly Notices of the Royal Astronomical Society} \textbf{\bibinfo{volume}{486}}, \bibinfo{pages}{1086–1093} (\bibinfo{year}{2019}), ISSN \bibinfo{issn}{1365-2966}, \urlprefix\url{http://dx.doi.org/10.1093/mnras/stz896}.

\bibitem[{\citenamefont{Thrane and Talbot}(2019)}]{Thrane_2019}
\bibinfo{author}{\bibfnamefont{E.}~\bibnamefont{Thrane}} \bibnamefont{and} \bibinfo{author}{\bibfnamefont{C.}~\bibnamefont{Talbot}}, \bibinfo{journal}{Publications of the Astronomical Society of Australia} \textbf{\bibinfo{volume}{36}} (\bibinfo{year}{2019}), ISSN \bibinfo{issn}{1448-6083}, \urlprefix\url{http://dx.doi.org/10.1017/pasa.2019.2}.

\bibitem[{\citenamefont{Vitale et~al.}(2021)\citenamefont{Vitale, Gerosa, Farr, and Taylor}}]{Vitale_2021}
\bibinfo{author}{\bibfnamefont{S.}~\bibnamefont{Vitale}}, \bibinfo{author}{\bibfnamefont{D.}~\bibnamefont{Gerosa}}, \bibinfo{author}{\bibfnamefont{W.~M.} \bibnamefont{Farr}}, \bibnamefont{and} \bibinfo{author}{\bibfnamefont{S.~R.} \bibnamefont{Taylor}}, \emph{\bibinfo{title}{Inferring the Properties of a Population of Compact Binaries in Presence of Selection Effects}} (\bibinfo{publisher}{Springer Singapore}, \bibinfo{year}{2021}), p. \bibinfo{pages}{1–60}, ISBN \bibinfo{isbn}{9789811547027}, \urlprefix\url{http://dx.doi.org/10.1007/978-981-15-4702-7_45-1}.

\bibitem[{\citenamefont{Fishbach et~al.}(2018)\citenamefont{Fishbach, Holz, and Farr}}]{Fishbach_2018}
\bibinfo{author}{\bibfnamefont{M.}~\bibnamefont{Fishbach}}, \bibinfo{author}{\bibfnamefont{D.~E.} \bibnamefont{Holz}}, \bibnamefont{and} \bibinfo{author}{\bibfnamefont{W.~M.} \bibnamefont{Farr}}, \bibinfo{journal}{The Astrophysical Journal Letters} \textbf{\bibinfo{volume}{863}}, \bibinfo{pages}{L41} (\bibinfo{year}{2018}), ISSN \bibinfo{issn}{2041-8213}, \urlprefix\url{http://dx.doi.org/10.3847/2041-8213/aad800}.

\bibitem[{\citenamefont{Abbott et~al.}(2023{\natexlab{b}})}]{KAGRA:2021duu}
\bibinfo{author}{\bibfnamefont{R.}~\bibnamefont{Abbott}} \bibnamefont{et~al.} (\bibinfo{collaboration}{KAGRA, VIRGO, LIGO Scientific}), \bibinfo{journal}{Phys. Rev. X} \textbf{\bibinfo{volume}{13}}, \bibinfo{pages}{011048} (\bibinfo{year}{2023}{\natexlab{b}}), \eprint{2111.03634}.

\bibitem[{\citenamefont{Farr et~al.}(2019)\citenamefont{Farr, Fishbach, Ye, and Holz}}]{Farr_2019}
\bibinfo{author}{\bibfnamefont{W.~M.} \bibnamefont{Farr}}, \bibinfo{author}{\bibfnamefont{M.}~\bibnamefont{Fishbach}}, \bibinfo{author}{\bibfnamefont{J.}~\bibnamefont{Ye}}, \bibnamefont{and} \bibinfo{author}{\bibfnamefont{D.~E.} \bibnamefont{Holz}}, \bibinfo{journal}{The Astrophysical Journal Letters} \textbf{\bibinfo{volume}{883}}, \bibinfo{pages}{L42} (\bibinfo{year}{2019}), \urlprefix\url{https://doi.org/10.3847%2F2041-8213%2Fab4284}.

\bibitem[{\citenamefont{Tiwari}(2018)}]{Tiwari_2018}
\bibinfo{author}{\bibfnamefont{V.}~\bibnamefont{Tiwari}}, \bibinfo{journal}{Classical and Quantum Gravity} \textbf{\bibinfo{volume}{35}}, \bibinfo{pages}{145009} (\bibinfo{year}{2018}), ISSN \bibinfo{issn}{1361-6382}, \urlprefix\url{http://dx.doi.org/10.1088/1361-6382/aac89d}.

\bibitem[{\citenamefont{Callister}(2021)}]{Callister:2021gxf}
\bibinfo{author}{\bibfnamefont{T.}~\bibnamefont{Callister}} (\bibinfo{year}{2021}), \eprint{2104.09508}.

\bibitem[{\citenamefont{Jeffreys}(1961)}]{Jefferys1961}
\bibinfo{author}{\bibfnamefont{H.}~\bibnamefont{Jeffreys}}, \emph{\bibinfo{title}{Theory of Probability}} (\bibinfo{publisher}{Oxford}, \bibinfo{year}{1961}), \bibinfo{edition}{3rd} ed.

\bibitem[{\citenamefont{Farah et~al.}(2024{\natexlab{b}})\citenamefont{Farah, Fishbach, and Holz}}]{farah:2023swu}
\bibinfo{author}{\bibfnamefont{A.~M.} \bibnamefont{Farah}}, \bibinfo{author}{\bibfnamefont{M.}~\bibnamefont{Fishbach}}, \bibnamefont{and} \bibinfo{author}{\bibfnamefont{D.~E.} \bibnamefont{Holz}}, \bibinfo{journal}{The Astrophysical Journal} \textbf{\bibinfo{volume}{962}}, \bibinfo{pages}{69} (\bibinfo{year}{2024}{\natexlab{b}}), \eprint{2308.05102}.

\bibitem[{\citenamefont{Abbott et~al.}(2023{\natexlab{c}})}]{KAGRA:2023pio}
\bibinfo{author}{\bibfnamefont{R.}~\bibnamefont{Abbott}} \bibnamefont{et~al.} (\bibinfo{collaboration}{KAGRA, VIRGO, LIGO Scientific}), \bibinfo{journal}{Astrophys. J. Suppl.} \textbf{\bibinfo{volume}{267}}, \bibinfo{pages}{29} (\bibinfo{year}{2023}{\natexlab{c}}), \eprint{2302.03676}.

\bibitem[{\citenamefont{Abbott et~al.}(2020{\natexlab{b}})}]{LIGOScientific:2020zkf}
\bibinfo{author}{\bibfnamefont{R.}~\bibnamefont{Abbott}} \bibnamefont{et~al.} (\bibinfo{collaboration}{LIGO Scientific, Virgo}), \bibinfo{journal}{Astrophys. J. Lett.} \textbf{\bibinfo{volume}{896}}, \bibinfo{pages}{L44} (\bibinfo{year}{2020}{\natexlab{b}}), \eprint{2006.12611}.

\bibitem[{\citenamefont{Moore and Gerosa}(2021)}]{Moore_2021}
\bibinfo{author}{\bibfnamefont{C.~J.} \bibnamefont{Moore}} \bibnamefont{and} \bibinfo{author}{\bibfnamefont{D.}~\bibnamefont{Gerosa}}, \bibinfo{journal}{Physical Review D} \textbf{\bibinfo{volume}{104}} (\bibinfo{year}{2021}), ISSN \bibinfo{issn}{2470-0029}, \urlprefix\url{http://dx.doi.org/10.1103/PhysRevD.104.083008}.

\bibitem[{\citenamefont{Talbot and Thrane}(2018)}]{Talbot:2018cva}
\bibinfo{author}{\bibfnamefont{C.}~\bibnamefont{Talbot}} \bibnamefont{and} \bibinfo{author}{\bibfnamefont{E.}~\bibnamefont{Thrane}}, \bibinfo{journal}{Astrophys. J.} \textbf{\bibinfo{volume}{856}}, \bibinfo{pages}{173} (\bibinfo{year}{2018}), \eprint{1801.02699}.

\bibitem[{\citenamefont{Fishbach et~al.}(2020)\citenamefont{Fishbach, Farr, and Holz}}]{Fishbach_2020}
\bibinfo{author}{\bibfnamefont{M.}~\bibnamefont{Fishbach}}, \bibinfo{author}{\bibfnamefont{W.~M.} \bibnamefont{Farr}}, \bibnamefont{and} \bibinfo{author}{\bibfnamefont{D.~E.} \bibnamefont{Holz}}, \bibinfo{journal}{The Astrophysical Journal Letters} \textbf{\bibinfo{volume}{891}}, \bibinfo{pages}{L31} (\bibinfo{year}{2020}), ISSN \bibinfo{issn}{2041-8213}, \urlprefix\url{http://dx.doi.org/10.3847/2041-8213/ab77c9}.

\bibitem[{\citenamefont{Abbott et~al.}(2016{\natexlab{b}})}]{KAGRA:2013rdx}
\bibinfo{author}{\bibfnamefont{B.~P.} \bibnamefont{Abbott}} \bibnamefont{et~al.} (\bibinfo{collaboration}{KAGRA, LIGO Scientific, Virgo}), \bibinfo{journal}{Living Rev. Rel.} \textbf{\bibinfo{volume}{19}}, \bibinfo{pages}{1} (\bibinfo{year}{2016}{\natexlab{b}}), \eprint{1304.0670}.

\bibitem[{\citenamefont{Gerosa et~al.}(2018)\citenamefont{Gerosa, Berti, O’Shaughnessy, Belczynski, Kesden, Wysocki, and Gladysz}}]{Gerosa_2018}
\bibinfo{author}{\bibfnamefont{D.}~\bibnamefont{Gerosa}}, \bibinfo{author}{\bibfnamefont{E.}~\bibnamefont{Berti}}, \bibinfo{author}{\bibfnamefont{R.}~\bibnamefont{O’Shaughnessy}}, \bibinfo{author}{\bibfnamefont{K.}~\bibnamefont{Belczynski}}, \bibinfo{author}{\bibfnamefont{M.}~\bibnamefont{Kesden}}, \bibinfo{author}{\bibfnamefont{D.}~\bibnamefont{Wysocki}}, \bibnamefont{and} \bibinfo{author}{\bibfnamefont{W.}~\bibnamefont{Gladysz}}, \bibinfo{journal}{Physical Review D} \textbf{\bibinfo{volume}{98}} (\bibinfo{year}{2018}), ISSN \bibinfo{issn}{2470-0029}, \urlprefix\url{http://dx.doi.org/10.1103/PhysRevD.98.084036}.

\bibitem[{\citenamefont{{Zaldarriaga} et~al.}(2018)\citenamefont{{Zaldarriaga}, {Kushnir}, and {Kollmeier}}}]{Zaldarriaga_2018}
\bibinfo{author}{\bibfnamefont{M.}~\bibnamefont{{Zaldarriaga}}}, \bibinfo{author}{\bibfnamefont{D.}~\bibnamefont{{Kushnir}}}, \bibnamefont{and} \bibinfo{author}{\bibfnamefont{J.~A.} \bibnamefont{{Kollmeier}}}, \bibinfo{journal}{Monthly Notices of the Royal Astronomical Society: Letters} \textbf{\bibinfo{volume}{473}}, \bibinfo{pages}{4174} (\bibinfo{year}{2018}), \eprint{1702.00885}.

\bibitem[{\citenamefont{Belczynski et~al.}(2020)\citenamefont{Belczynski, Klencki, Fields, Olejak, Berti, Meynet, Fryer, Holz, O’Shaughnessy, Brown et~al.}}]{Belczynski_2020}
\bibinfo{author}{\bibfnamefont{K.}~\bibnamefont{Belczynski}}, \bibinfo{author}{\bibfnamefont{J.}~\bibnamefont{Klencki}}, \bibinfo{author}{\bibfnamefont{C.~E.} \bibnamefont{Fields}}, \bibinfo{author}{\bibfnamefont{A.}~\bibnamefont{Olejak}}, \bibinfo{author}{\bibfnamefont{E.}~\bibnamefont{Berti}}, \bibinfo{author}{\bibfnamefont{G.}~\bibnamefont{Meynet}}, \bibinfo{author}{\bibfnamefont{C.~L.} \bibnamefont{Fryer}}, \bibinfo{author}{\bibfnamefont{D.~E.} \bibnamefont{Holz}}, \bibinfo{author}{\bibfnamefont{R.}~\bibnamefont{O’Shaughnessy}}, \bibinfo{author}{\bibfnamefont{D.~A.} \bibnamefont{Brown}}, \bibnamefont{et~al.}, \bibinfo{journal}{Astronomy and; Astrophysics} \textbf{\bibinfo{volume}{636}}, \bibinfo{pages}{A104} (\bibinfo{year}{2020}), ISSN \bibinfo{issn}{1432-0746}, \urlprefix\url{http://dx.doi.org/10.1051/0004-6361/201936528}.

\bibitem[{\citenamefont{{Gerosa} and {Berti}}(2017)}]{Gerosa_2017}
\bibinfo{author}{\bibfnamefont{D.}~\bibnamefont{{Gerosa}}} \bibnamefont{and} \bibinfo{author}{\bibfnamefont{E.}~\bibnamefont{{Berti}}}, \bibinfo{journal}{Physical Review D} \textbf{\bibinfo{volume}{95}}, \bibinfo{eid}{124046} (\bibinfo{year}{2017}), \eprint{1703.06223}.

\bibitem[{\citenamefont{{van Son} et~al.}(2022)\citenamefont{{van Son}, {de Mink}, {Callister}, {Justham}, {Renzo}, {Wagg}, {Broekgaarden}, {Kummer}, {Pakmor}, and {Mandel}}}]{2022ApJ...931...17V}
\bibinfo{author}{\bibfnamefont{L.~A.~C.} \bibnamefont{{van Son}}}, \bibinfo{author}{\bibfnamefont{S.~E.} \bibnamefont{{de Mink}}}, \bibinfo{author}{\bibfnamefont{T.}~\bibnamefont{{Callister}}}, \bibinfo{author}{\bibfnamefont{S.}~\bibnamefont{{Justham}}}, \bibinfo{author}{\bibfnamefont{M.}~\bibnamefont{{Renzo}}}, \bibinfo{author}{\bibfnamefont{T.}~\bibnamefont{{Wagg}}}, \bibinfo{author}{\bibfnamefont{F.~S.} \bibnamefont{{Broekgaarden}}}, \bibinfo{author}{\bibfnamefont{F.}~\bibnamefont{{Kummer}}}, \bibinfo{author}{\bibfnamefont{R.}~\bibnamefont{{Pakmor}}}, \bibnamefont{and} \bibinfo{author}{\bibfnamefont{I.}~\bibnamefont{{Mandel}}}, \bibinfo{journal}{The Astrophysical Journal} \textbf{\bibinfo{volume}{931}}, \bibinfo{eid}{17} (\bibinfo{year}{2022}), \eprint{2110.01634}.

\bibitem[{\citenamefont{Godfrey et~al.}(2024)\citenamefont{Godfrey, Edelman, and Farr}}]{godfrey2024cosmiccousinsidentificationsubpopulation}
\bibinfo{author}{\bibfnamefont{J.}~\bibnamefont{Godfrey}}, \bibinfo{author}{\bibfnamefont{B.}~\bibnamefont{Edelman}}, \bibnamefont{and} \bibinfo{author}{\bibfnamefont{B.}~\bibnamefont{Farr}}, \emph{\bibinfo{title}{Cosmic cousins: Identification of a subpopulation of binary black holes consistent with isolated binary evolution}} (\bibinfo{year}{2024}), \eprint{2304.01288}, \urlprefix\url{https://arxiv.org/abs/2304.01288}.

\bibitem[{\citenamefont{Li et~al.}(2024)\citenamefont{Li, Wang, Tang, and Fan}}]{li2024}
\bibinfo{author}{\bibfnamefont{Y.-J.} \bibnamefont{Li}}, \bibinfo{author}{\bibfnamefont{Y.-Z.} \bibnamefont{Wang}}, \bibinfo{author}{\bibfnamefont{S.-P.} \bibnamefont{Tang}}, \bibnamefont{and} \bibinfo{author}{\bibfnamefont{Y.-Z.} \bibnamefont{Fan}}, \emph{\bibinfo{title}{Resolving the stellar-collapse and hierarchical-merger origins of the coalescing black holes}} (\bibinfo{year}{2024}), \eprint{2303.02973}, \urlprefix\url{https://arxiv.org/abs/2303.02973}.

\bibitem[{\citenamefont{{Bavera} et~al.}(2020)\citenamefont{{Bavera}, {Fragos}, {Qin}, {Zapartas}, {Neijssel}, {Mandel}, {Batta}, {Gaebel}, {Kimball}, and {Stevenson}}}]{bavera2020}
\bibinfo{author}{\bibfnamefont{S.~S.} \bibnamefont{{Bavera}}}, \bibinfo{author}{\bibfnamefont{T.}~\bibnamefont{{Fragos}}}, \bibinfo{author}{\bibfnamefont{Y.}~\bibnamefont{{Qin}}}, \bibinfo{author}{\bibfnamefont{E.}~\bibnamefont{{Zapartas}}}, \bibinfo{author}{\bibfnamefont{C.~J.} \bibnamefont{{Neijssel}}}, \bibinfo{author}{\bibfnamefont{I.}~\bibnamefont{{Mandel}}}, \bibinfo{author}{\bibfnamefont{A.}~\bibnamefont{{Batta}}}, \bibinfo{author}{\bibfnamefont{S.~M.} \bibnamefont{{Gaebel}}}, \bibinfo{author}{\bibfnamefont{C.}~\bibnamefont{{Kimball}}}, \bibnamefont{and} \bibinfo{author}{\bibfnamefont{S.}~\bibnamefont{{Stevenson}}}, \bibinfo{journal}{Astronomy and; Astrophysics} \textbf{\bibinfo{volume}{635}}, \bibinfo{eid}{A97} (\bibinfo{year}{2020}), \eprint{1906.12257}.

\bibitem[{\citenamefont{{Kalogera}}(1996)}]{kalogera_1996}
\bibinfo{author}{\bibfnamefont{V.}~\bibnamefont{{Kalogera}}}, \bibinfo{journal}{The Astrophysical Journal} \textbf{\bibinfo{volume}{471}}, \bibinfo{pages}{352} (\bibinfo{year}{1996}), \eprint{astro-ph/9605186}.

\bibitem[{\citenamefont{Farmer et~al.}(2019)\citenamefont{Farmer, Renzo, de~Mink, Marchant, and Justham}}]{Farmer_2019}
\bibinfo{author}{\bibfnamefont{R.}~\bibnamefont{Farmer}}, \bibinfo{author}{\bibfnamefont{M.}~\bibnamefont{Renzo}}, \bibinfo{author}{\bibfnamefont{S.~E.} \bibnamefont{de~Mink}}, \bibinfo{author}{\bibfnamefont{P.}~\bibnamefont{Marchant}}, \bibnamefont{and} \bibinfo{author}{\bibfnamefont{S.}~\bibnamefont{Justham}}, \bibinfo{journal}{The Astrophysical Journal} \textbf{\bibinfo{volume}{887}}, \bibinfo{pages}{53} (\bibinfo{year}{2019}), ISSN \bibinfo{issn}{1538-4357}, \urlprefix\url{http://dx.doi.org/10.3847/1538-4357/ab518b}.

\bibitem[{\citenamefont{Farag et~al.}(2022)\citenamefont{Farag, Renzo, Farmer, Chidester, and Timmes}}]{Farag:2022jcc}
\bibinfo{author}{\bibfnamefont{E.}~\bibnamefont{Farag}}, \bibinfo{author}{\bibfnamefont{M.}~\bibnamefont{Renzo}}, \bibinfo{author}{\bibfnamefont{R.}~\bibnamefont{Farmer}}, \bibinfo{author}{\bibfnamefont{M.~T.} \bibnamefont{Chidester}}, \bibnamefont{and} \bibinfo{author}{\bibfnamefont{F.~X.} \bibnamefont{Timmes}}, \bibinfo{journal}{Astrophys. J.} \textbf{\bibinfo{volume}{937}}, \bibinfo{pages}{112} (\bibinfo{year}{2022}), \eprint{2208.09624}.

\bibitem[{\citenamefont{{Broekgaarden} et~al.}(2022)\citenamefont{{Broekgaarden}, {Berger}, {Stevenson}, {Justham}, {Mandel}, {Chru{\'s}li{\'n}ska}, {van Son}, {Wagg}, {Vigna-G{\'o}mez}, {de Mink} et~al.}}]{2022MNRAS.516.5737B}
\bibinfo{author}{\bibfnamefont{F.~S.} \bibnamefont{{Broekgaarden}}}, \bibinfo{author}{\bibfnamefont{E.}~\bibnamefont{{Berger}}}, \bibinfo{author}{\bibfnamefont{S.}~\bibnamefont{{Stevenson}}}, \bibinfo{author}{\bibfnamefont{S.}~\bibnamefont{{Justham}}}, \bibinfo{author}{\bibfnamefont{I.}~\bibnamefont{{Mandel}}}, \bibinfo{author}{\bibfnamefont{M.}~\bibnamefont{{Chru{\'s}li{\'n}ska}}}, \bibinfo{author}{\bibfnamefont{L.~A.~C.} \bibnamefont{{van Son}}}, \bibinfo{author}{\bibfnamefont{T.}~\bibnamefont{{Wagg}}}, \bibinfo{author}{\bibfnamefont{A.}~\bibnamefont{{Vigna-G{\'o}mez}}}, \bibinfo{author}{\bibfnamefont{S.~E.} \bibnamefont{{de Mink}}}, \bibnamefont{et~al.}, \bibinfo{journal}{Monthly Notices of the Royal Astronomical Society} \textbf{\bibinfo{volume}{516}}, \bibinfo{pages}{5737} (\bibinfo{year}{2022}), \eprint{2112.05763}.

\bibitem[{\citenamefont{{van Son} et~al.}(2023)\citenamefont{{van Son}, {de Mink}, {Chru{\'s}li{\'n}ska}, {Conroy}, {Pakmor}, and {Hernquist}}}]{2023ApJ...948..105V}
\bibinfo{author}{\bibfnamefont{L.~A.~C.} \bibnamefont{{van Son}}}, \bibinfo{author}{\bibfnamefont{S.~E.} \bibnamefont{{de Mink}}}, \bibinfo{author}{\bibfnamefont{M.}~\bibnamefont{{Chru{\'s}li{\'n}ska}}}, \bibinfo{author}{\bibfnamefont{C.}~\bibnamefont{{Conroy}}}, \bibinfo{author}{\bibfnamefont{R.}~\bibnamefont{{Pakmor}}}, \bibnamefont{and} \bibinfo{author}{\bibfnamefont{L.}~\bibnamefont{{Hernquist}}}, \bibinfo{journal}{The Astrophysical Journal} \textbf{\bibinfo{volume}{948}}, \bibinfo{eid}{105} (\bibinfo{year}{2023}), \eprint{2209.03385}.

\bibitem[{\citenamefont{Mapelli et~al.}(2022)\citenamefont{Mapelli, Bouffanais, Santoliquido, Arca Sedda, and Artale}}]{Mapelli_2022}
\bibinfo{author}{\bibfnamefont{M.}~\bibnamefont{Mapelli}}, \bibinfo{author}{\bibfnamefont{Y.}~\bibnamefont{Bouffanais}}, \bibinfo{author}{\bibfnamefont{F.}~\bibnamefont{Santoliquido}}, \bibinfo{author}{\bibfnamefont{M.}~\bibnamefont{Arca Sedda}}, \bibnamefont{and} \bibinfo{author}{\bibfnamefont{M.~C.} \bibnamefont{Artale}}, \bibinfo{journal}{Monthly Notices of the Royal Astronomical Society} \textbf{\bibinfo{volume}{511}}, \bibinfo{pages}{5797–5816} (\bibinfo{year}{2022}), ISSN \bibinfo{issn}{1365-2966}, \urlprefix\url{http://dx.doi.org/10.1093/mnras/stac422}.

\bibitem[{\citenamefont{Biscoveanu et~al.}(2022)\citenamefont{Biscoveanu, Callister, Haster, Ng, Vitale, and Farr}}]{Biscoveanu_2022}
\bibinfo{author}{\bibfnamefont{S.}~\bibnamefont{Biscoveanu}}, \bibinfo{author}{\bibfnamefont{T.~A.} \bibnamefont{Callister}}, \bibinfo{author}{\bibfnamefont{C.-J.} \bibnamefont{Haster}}, \bibinfo{author}{\bibfnamefont{K.~K.~Y.} \bibnamefont{Ng}}, \bibinfo{author}{\bibfnamefont{S.}~\bibnamefont{Vitale}}, \bibnamefont{and} \bibinfo{author}{\bibfnamefont{W.~M.} \bibnamefont{Farr}}, \bibinfo{journal}{The Astrophysical Journal Letters} \textbf{\bibinfo{volume}{932}}, \bibinfo{pages}{L19} (\bibinfo{year}{2022}), ISSN \bibinfo{issn}{2041-8213}, \urlprefix\url{http://dx.doi.org/10.3847/2041-8213/ac71a8}.

\bibitem[{\citenamefont{{Heinzel} et~al.}(2024{\natexlab{a}})\citenamefont{{Heinzel}, {Vitale}, and {Biscoveanu}}}]{2024PhRvD.109j3006H}
\bibinfo{author}{\bibfnamefont{J.}~\bibnamefont{{Heinzel}}}, \bibinfo{author}{\bibfnamefont{S.}~\bibnamefont{{Vitale}}}, \bibnamefont{and} \bibinfo{author}{\bibfnamefont{S.}~\bibnamefont{{Biscoveanu}}}, \bibinfo{journal}{Physical Review D} \textbf{\bibinfo{volume}{109}}, \bibinfo{eid}{103006} (\bibinfo{year}{2024}{\natexlab{a}}), \eprint{2312.00993}.

\bibitem[{\citenamefont{{Rinaldi} et~al.}(2024)\citenamefont{{Rinaldi}, {Del Pozzo}, {Mapelli}, {Lorenzo-Medina}, and {Dent}}}]{rinaldi_2024}
\bibinfo{author}{\bibfnamefont{S.}~\bibnamefont{{Rinaldi}}}, \bibinfo{author}{\bibfnamefont{W.}~\bibnamefont{{Del Pozzo}}}, \bibinfo{author}{\bibfnamefont{M.}~\bibnamefont{{Mapelli}}}, \bibinfo{author}{\bibfnamefont{A.}~\bibnamefont{{Lorenzo-Medina}}}, \bibnamefont{and} \bibinfo{author}{\bibfnamefont{T.}~\bibnamefont{{Dent}}}, \bibinfo{journal}{Astronomy and; Astrophysics} \textbf{\bibinfo{volume}{684}}, \bibinfo{eid}{A204} (\bibinfo{year}{2024}), \eprint{2310.03074}.

\bibitem[{\citenamefont{{Heinzel} et~al.}(2024{\natexlab{b}})\citenamefont{{Heinzel}, {Mould}, and {Vitale}}}]{Heinzel_2024}
\bibinfo{author}{\bibfnamefont{J.}~\bibnamefont{{Heinzel}}}, \bibinfo{author}{\bibfnamefont{M.}~\bibnamefont{{Mould}}}, \bibnamefont{and} \bibinfo{author}{\bibfnamefont{S.}~\bibnamefont{{Vitale}}}, \bibinfo{journal}{arXiv e-prints} \bibinfo{eid}{arXiv:2406.16844} (\bibinfo{year}{2024}{\natexlab{b}}), \eprint{2406.16844}.

\bibitem[{\citenamefont{Baibhav et~al.}(2020)\citenamefont{Baibhav, Gerosa, Berti, Wong, Helfer, and Mould}}]{Baibhav_2020}
\bibinfo{author}{\bibfnamefont{V.}~\bibnamefont{Baibhav}}, \bibinfo{author}{\bibfnamefont{D.}~\bibnamefont{Gerosa}}, \bibinfo{author}{\bibfnamefont{E.}~\bibnamefont{Berti}}, \bibinfo{author}{\bibfnamefont{K.~W.~K.} \bibnamefont{Wong}}, \bibinfo{author}{\bibfnamefont{T.}~\bibnamefont{Helfer}}, \bibnamefont{and} \bibinfo{author}{\bibfnamefont{M.}~\bibnamefont{Mould}}, \bibinfo{journal}{Physical Review D} \textbf{\bibinfo{volume}{102}} (\bibinfo{year}{2020}), ISSN \bibinfo{issn}{2470-0029}, \urlprefix\url{http://dx.doi.org/10.1103/PhysRevD.102.043002}.

\bibitem[{\citenamefont{Fishbach et~al.}(2022)\citenamefont{Fishbach, Kimball, and Kalogera}}]{Fishbach_2022}
\bibinfo{author}{\bibfnamefont{M.}~\bibnamefont{Fishbach}}, \bibinfo{author}{\bibfnamefont{C.}~\bibnamefont{Kimball}}, \bibnamefont{and} \bibinfo{author}{\bibfnamefont{V.}~\bibnamefont{Kalogera}}, \bibinfo{journal}{The Astrophysical Journal Letters} \textbf{\bibinfo{volume}{935}}, \bibinfo{pages}{L26} (\bibinfo{year}{2022}), ISSN \bibinfo{issn}{2041-8213}, \urlprefix\url{http://dx.doi.org/10.3847/2041-8213/ac86c4}.

\bibitem[{\citenamefont{{Kremer} et~al.}(2020)\citenamefont{{Kremer}, {Spera}, {Becker}, {Chatterjee}, {Di Carlo}, {Fragione}, {Rodriguez}, {Ye}, and {Rasio}}}]{Kremer2020}
\bibinfo{author}{\bibfnamefont{K.}~\bibnamefont{{Kremer}}}, \bibinfo{author}{\bibfnamefont{M.}~\bibnamefont{{Spera}}}, \bibinfo{author}{\bibfnamefont{D.}~\bibnamefont{{Becker}}}, \bibinfo{author}{\bibfnamefont{S.}~\bibnamefont{{Chatterjee}}}, \bibinfo{author}{\bibfnamefont{U.~N.} \bibnamefont{{Di Carlo}}}, \bibinfo{author}{\bibfnamefont{G.}~\bibnamefont{{Fragione}}}, \bibinfo{author}{\bibfnamefont{C.~L.} \bibnamefont{{Rodriguez}}}, \bibinfo{author}{\bibfnamefont{C.~S.} \bibnamefont{{Ye}}}, \bibnamefont{and} \bibinfo{author}{\bibfnamefont{F.~A.} \bibnamefont{{Rasio}}}, \bibinfo{journal}{The Astrophysical Journal} \textbf{\bibinfo{volume}{903}}, \bibinfo{eid}{45} (\bibinfo{year}{2020}), \eprint{2006.10771}.

\bibitem[{\citenamefont{Ezquiaga and Holz}(2022)}]{spectral_sirens}
\bibinfo{author}{\bibfnamefont{J.~M.} \bibnamefont{Ezquiaga}} \bibnamefont{and} \bibinfo{author}{\bibfnamefont{D.~E.} \bibnamefont{Holz}}, \bibinfo{journal}{Phys. Rev. Lett.} \textbf{\bibinfo{volume}{129}}, \bibinfo{pages}{061102} (\bibinfo{year}{2022}), \eprint{2202.08240}.

\end{thebibliography}

\newpage
\appendix
\onecolumngrid
\section{Full Posterior distributions }\label{app:appendix}

\begin{figure}[!htb]
\begin{center}
\includegraphics[width=\textwidth]{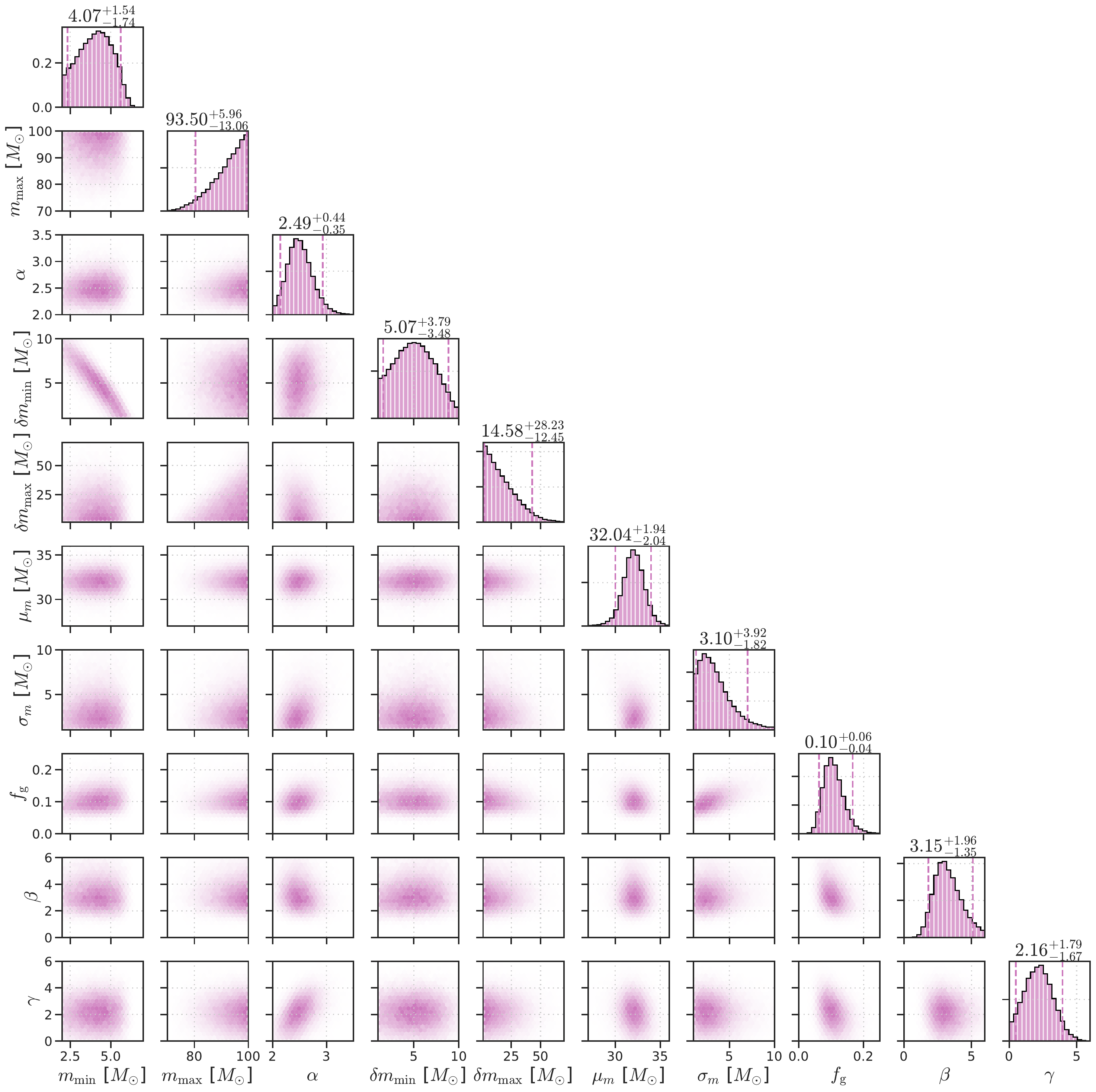}
\caption{\label{fig:plbcorner} Corner plot showing the full posterior distribution for the \plpeak model.}
\end{center}
\end{figure}

\begin{figure}[!htb]
\begin{center}
\includegraphics[width=\textwidth]{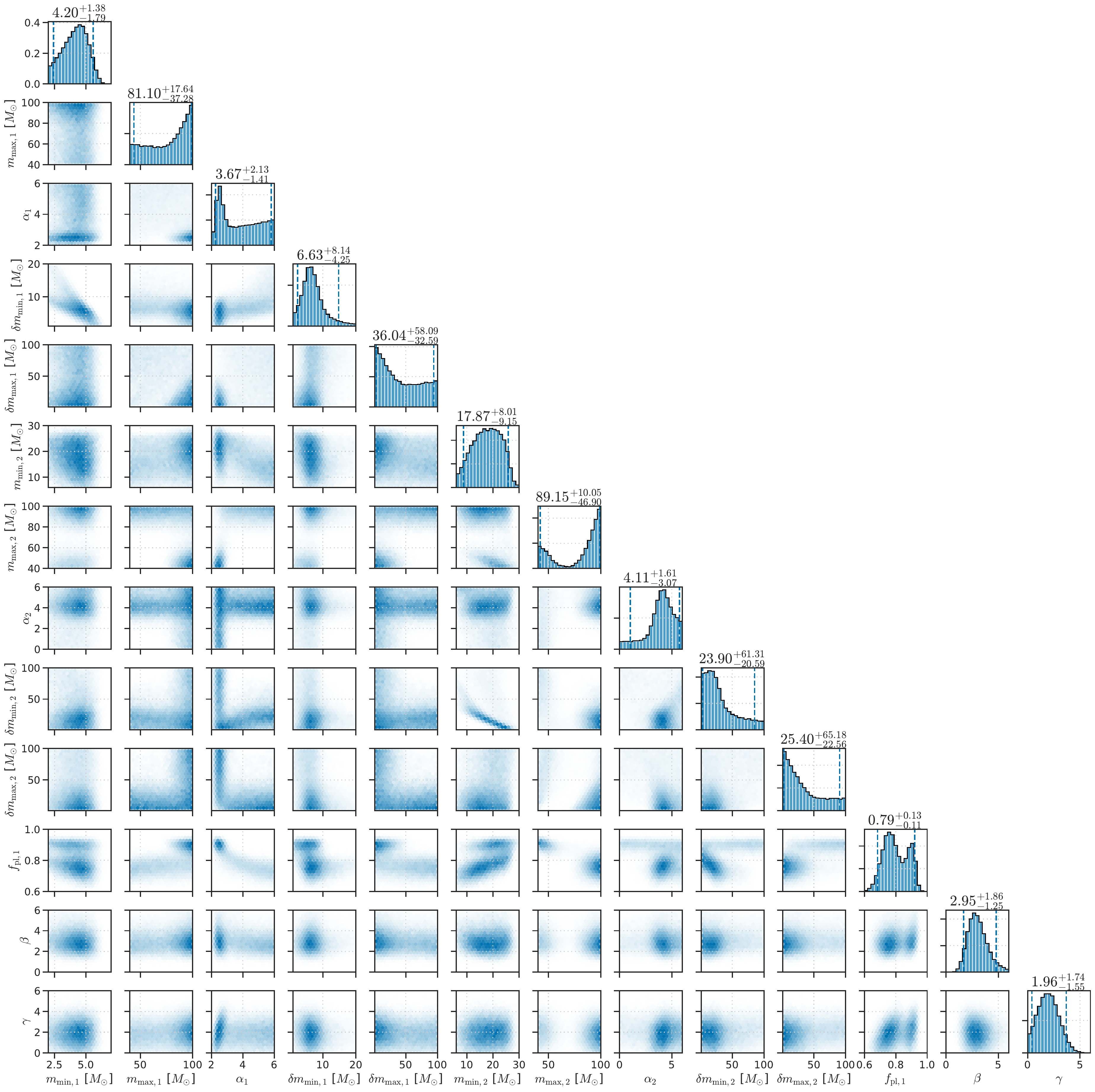}
\caption{\label{fig:plplcorner} Corner plot showing the full posterior distribution for the \plpl model.}
\end{center}
\end{figure}

\begin{figure}[!htb]
\begin{center}
\includegraphics[width=\textwidth]{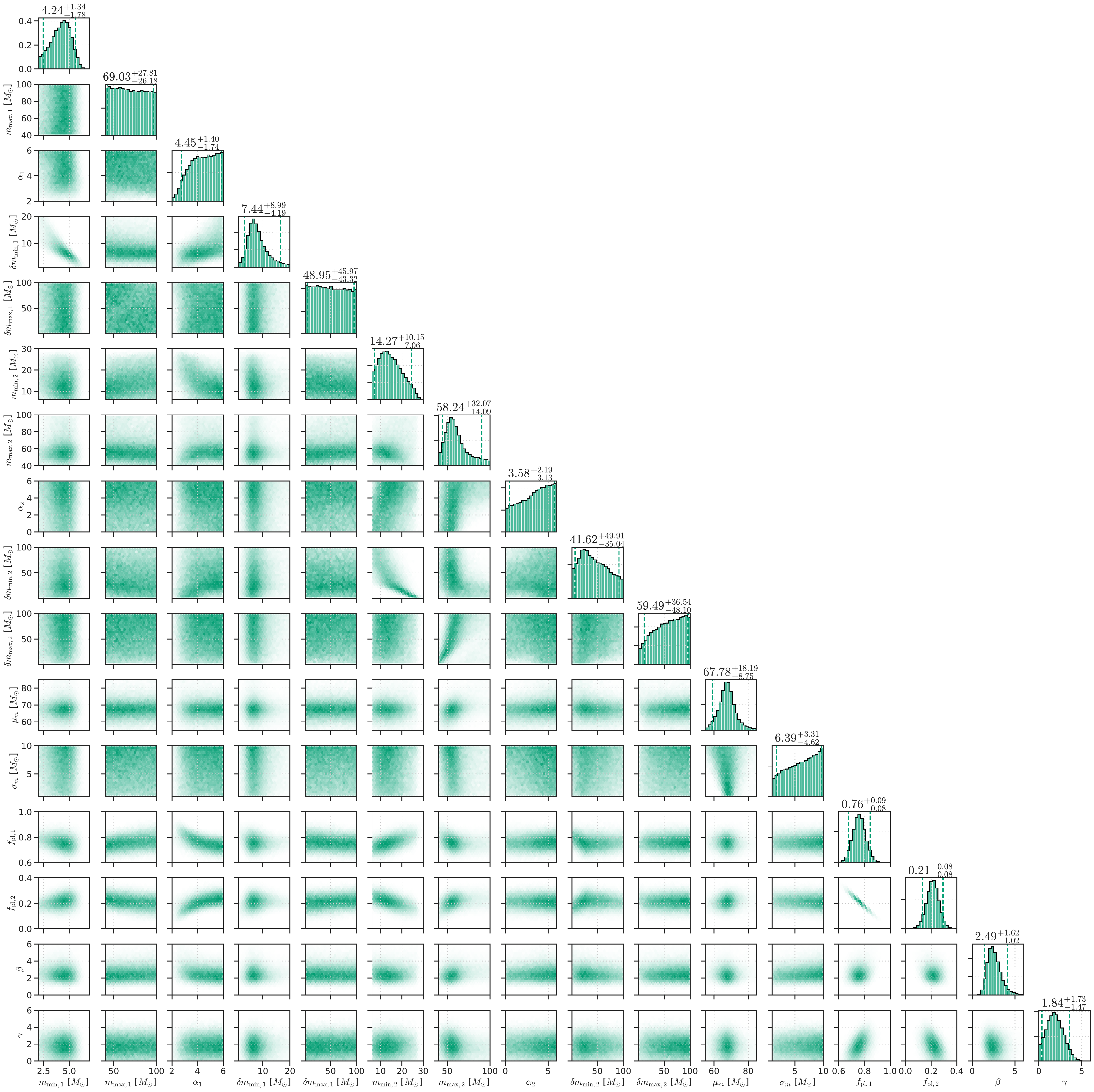}
\caption{\label{fig:plplbcorner} Corner plot showing the full posterior distribution for the \plplpeak model.}
\end{center}
\end{figure}

\end{document}